\documentclass[aps,prd,preprint,nofootinbib,superscriptaddress]{revtex4}
\usepackage{color}
\usepackage{graphicx}
\usepackage{psfrag}

\begin{document}
\preprint{UTHEP-626}
\preprint{OU-HET-700-2011}

\title{Interpolation between the $\epsilon$ and $p$ regimes}

\newcommand{\TsukubaA}{
  Graduate School of Pure and Applied Sciences, University of Tsukuba,
  Tsukuba 305-8571, Japan
}
\newcommand{\TsukubaB}{
  Center for Computational Sciences, University of Tsukuba,
  Tsukuba 305-8577, Japan
}

\newcommand{\Osaka}{
  Department of Physics, Osaka University,
  Toyonaka 560-0043, Japan
}

\author{Sinya~Aoki}
\affiliation{\TsukubaA}
\affiliation{\TsukubaB}

\author{Hidenori~Fukaya}
\affiliation{\Osaka}

\begin{abstract}
We reconsider chiral perturbation theory in a finite volume and 
develop a new computational scheme which smoothly
interpolates the conventional $\epsilon$ and $p$ regimes.
The counting rule is kept essentially the same as in the $p$ expansion.
The zero-momentum modes of Nambu-Goldstone bosons 
are, however, treated separately and partly integrated out to all orders
as in the $\epsilon$ expansion.  
In this new scheme, the theory remains infra-red finite
even in the chiral limit, while 
the chiral-logarithmic effects are kept present.
We calculate the two-point function in the pseudoscalar channel
and show that the correlator 
has a constant contribution in addition to the conventional $\cosh$ function
of time $t$.
This constant term rapidly disappears in the $p$ regime
but it is indispensable for a smooth convergence of the formula 
to the $\epsilon$ regime result.
Our calculation is useful to precisely estimate the finite volume effects in
lattice QCD simulations on the pion mass $M_\pi$ and kaon mass $M_K$, 
as well as their decay constants $F_\pi$ and $F_K$. 
\end{abstract}

\maketitle

\section{Introduction}
\label{sec:intro}
\setcounter{equation}{0}

Recent progress in lattice QCD has made it possible
to simulate QCD in a realistic set-up, {\it i.e.}
with the (2+1)-flavor sea quark masses 
near the physical point.
As the precision of the data analysis goes high,
however, more precise study of systematic effects 
is required.
Finite volume effects are particularly important
when quark masses are reduced to near the chiral limit,
since the correlation length of the system rapidly grows,
which is induced by the dynamical chiral symmetry breaking
\cite{Nambu:1961tp}.

The chiral symmetry breaking
makes a mass gap between the Nambu-Goldstone
bosons, which eventually become massless in the chiral limit,
and the other hadrons, which retain a mass around
the QCD scale $\Lambda_{\rm QCD}$.
It is, therefore, the pions that are the most responsible
for the effects of the finite volume $V$ when 
the size of the system $L$ or $V^{1/4}$ 
is well above $1/\Lambda_{\rm QCD}$.

With this motivation, a number of studies have been devoted to
understand the finite volume effects within 
the theory of pions, which is known as
chiral perturbation theory (ChPT) 
\cite{Weinberg:1968de, Gasser:1983yg}.
Using the lattice data for the low-energy constants
as inputs, one can quantify the 
finite volume effects from the pion fields.
These studies are also useful for improving the
determination of the input low-energy constants themselves.

To investigate ChPT in a finite volume, 
two perturbative approaches have 
been proposed so far. 
One is the $p$ expansion 
\cite{Gasser:1986vb, Gasser:1987zq, Bernard:2001yj, Colangelo:2005gd}, 
which has just the same form as the perturbative series in an infinite volume,
but momentum integration is performed in a 
discrete space in the units of $1/L$. 
Denoting the mass of a generic (pseudo)
Nambu-Goldstone boson by $M$, this $p$ expansion is valid when $ML \gg 1$,
which is called the $p$ regime.

A nonperturbative technique is required when $ML \ll 1$ (the $\epsilon$ regime)
since the zero-mode's contribution to the 
propagator of the pseudo Nambu-Goldstone bosons
blows up and fluctuation $\sim 1/M^2$ cannot be perturbatively treated, 
which is well-known as the critical fluctuation
due to the symmetry breaking.
A solution to this problem was given in terms 
of the so-called $\epsilon$ expansion in
Refs.~\cite{Gasser:1987ah, Neuberger:1987zz, 
Hansen:1990un, Hasenfratz:1989pk, Leutwyler:1992yt} 
and later the study is extended in various directions
\cite{Damgaard:2001js, Damgaard:2002qe, Hernandez:2002ds, Damgaard:2007ep, 
Bernardoni:2007hi, Bernardoni:2008ei, Akemann:2008vp, Shindler:2009ri, 
Bar:2008th, Bernardoni:2009sx, Bar:2010zj, Bernardoni:2010nf, Lehner:2010mv}.
In this scheme the zero-momentum mode is
separately treated and integrated out exactly, 
while all the remaining non-zero momentum
modes are treated perturbatively. 
Since the $\epsilon$ expansion treats the mass term 
as a next-to-leading order (NLO) contribution,
the number of terms in the chiral Lagrangian is reduced compared to the 
$p$ expansion and
the typical {\it chiral-logs} are invisible 
in the calculation at NLO.
Note here that the {\it exact} integration here refers to 
the term that is leading order in the quark masses $m$. 

One may ask what happens in between: when $ML\sim 1$. 
The answer should be given in either ways of the expansions since
the $p$ and $\epsilon$ expansions should eventually converge to 
give the same result as the order of loop expansion increases.
But it is difficult already at the two-loop level,
to confirm such a convergence between the $p$ regime \cite{Colangelo:2005gd}
and $\epsilon$ regime \cite{Lehner:2010mv} calculations
unless one directly checks the numerical values,
since their analytic forms look quite different.
It is, therefore, important and useful for the practical calculation, 
to find a new way of expansion which smoothly interpolates 
the $p$ and $\epsilon$ expansions while keeping
the calculation at the one-loop level.
Intuitively, this one-loop level interpolation should be possible
in the simplest way, by keeping all the terms that appear
in the NLO Lagrangian in both expansions.

In fact, such a calculation is demanding. 
Although recent developments in computational facilities
have allowed us to simulate unquenched lattice QCD near the chiral limit, 
it is still difficult to fully satisfy the condition $M L \gg 1$.
On the other hand, no study has until now reached 
deep inside the $\epsilon$ regime keeping $ML \ll 1$ 
\cite{DeGrand:2006nv, Lang:2006ab, Hasenfratz:2007yj, 
Fukaya:2007pn, Hasenfratz:2008fg, Hasenfratz:2008ce, Bar:2009qa, Jansen:2009tt}.
Although results have often been compared favorably to 
the $\epsilon$ expansion of ChPT,
there may still be large systematic errors due to 
the condition $M L \ll 1$ not being well fulfilled.

Recently a new approach which smoothly 
connects the $p$ expansion and $\epsilon$ expansion
(and which remains valid even in the region $M L\sim 1$)
was proposed in Ref.~\cite{Damgaard:2008zs}.
The new prescription is to keep the counting rule of the $p$ expansion
but treat the zero-mode non-perturbatively as
in the $\epsilon$ expansion.
This new expansion was applied to the 
calculation of the chiral condensate 
(and the spectral density of the Dirac operator) 
to NLO and successful in maintaining the features
of the both regimes: non-perturbative behavior 
of the zero-modes and chiral logarithms.
The results are kept infra-red (IR) finite even in the 
chiral limit \cite{Damgaard:1997ye, Wilke:1997gf, Akemann:1998ta}
and show a good convergence
to the conventional result \cite{Smilga:1993in, Osborn:1998qb}
in the $p$ expansion for the large (valence) quark mass region.
A good agreement with a lattice QCD calculation was reported in
Refs.~\cite{Fukaya:2009fh, Fukaya:2010na}.

In this paper, we extend the calculation of Ref.~\cite{Damgaard:2008zs}
to the two-point functions in the pseudoscalar channel.
We find that 
the correlator is expressed by a simple hyperbolic cosine function of time $t$
plus an additional constant term, which smoothly connects
the conventional $p$ regime results and those in the $\epsilon$ regime. 
The constant contribution is a peculiar feature
of the $\epsilon$ expansion. We find that this constant is indispensable 
to keep the correlator IR finite, and show how and where it becomes
negligible as entering the $p$ expansion regime. 
Our results are useful to precisely estimate the finite volume effects
in lattice QCD 
on the pion mass $M_\pi$ and kaon mass $M_K$, 
as well as their decay constants $F_\pi$ and $F_K$.

The rest of our paper is organized as follows.
In Section \ref{sec:counting}, we describe in detail our new perturbative 
counting rule in ChPT and the computation scheme 
which consists of three steps. 
For the first step, the chiral Lagrangian 
in terms of non-self-contracting (NSC) vertices 
(whose definition is given in the following sections) of non-zero momentum modes
is calculated in Section  \ref{sec:Leff}.
The second step is to collect the one-loop diagrams of the correlator
and perform the non-zero mode's perturbative integrals (Section \ref{sec:xi}).
The final step is non-perturbative zero-mode's 
integration in Section \ref{sec:U0}.
The results for the two-point functions in the theory with a general number of flavors
are presented in Section \ref{sec:2pt} (see Eq.(\ref{eq:PPd3x})).
For more practical uses, explicit formulas for the $N_f=2$ and 2+1 cases
are given in Section \ref{sec:example} (see Eq.(\ref{eq:pion})) as well as 
how to compare the results with the lattice QCD data.
Our calculation suggests that there exists a simplified 
short-cut prescription which reproduces the same results. 
We discuss this simplified scheme in \ref{sec:discussion}.
Conclusions are given in Section \ref{sec:conclusion}.
\section{New  chiral expansion at finite volume}
\label{sec:counting}

In this section we review the new counting rule
of chiral perturbation which was first proposed by 
Ref.~\cite{Damgaard:2008zs}. We also present our strategy for
the calculation of two-point functions.

We consider an $N_f$-flavor chiral Lagrangian 
in a finite volume ($V=L^3T$),
\begin{eqnarray}
\mathcal{L}
&=&
\frac{F^2}{4}{\rm Tr}
[\partial_\mu U(x)^\dagger\partial_\mu U(x) ]
-\frac{\Sigma}{2}{\rm Tr}
[\mathcal{M}^{\dagger}e^{i\theta /N_f}U(x)
+U(x)^\dagger e^{-i\theta /N_f}\mathcal{M}]+\cdots,
\end{eqnarray}
where $U(x)\in SU(N_f)$ and $\theta$ denotes 
the vacuum angle, while $\Sigma$ is the chiral condensate
and $F$ denotes the pion decay constant both in the chiral limit.
We note that the higher order terms are not explicitly shown here but
exist, which is indicated by ellipses.

In the partially quenched case, we use the replica method
where the calculations are done within an 
$(N_f+N_v+(N-N_v))$-flavor theory and 
the limit $N \to 0$ is taken
\cite{Damgaard:1999ic, Damgaard:2000gh, Damgaard:2000di}
\footnote{We do not consider the fully quenched theory in this
work. We thus have $N_f > 0$ in all that follows.}.
Physical unquenched $N_f$-flavor theory results 
can be obtained by
simply taking $m_v=m_f$ where $m_f$ is
one of the physical quark masses.

For the mass matrix, we thus consider a general non-degenerate form:
\begin{eqnarray}
\mathcal{M} &=& 
\mathrm{diag}(\underbrace{m_{v_1},\cdots}_{N_{1}} 
\underbrace{m_{v_2}\cdots}_{N_{2}},
\underbrace{m_u,m_d,m_s,\cdots}_{N_f}).
\end{eqnarray}
where we have $N=N_{1}+N_{2}$ replica flavors and $N_f$ physical flavors. 
Since our target is a single meson system which consists of two quarks,
we have written the valence part as if there were
two different sets of degenerate flavors, 
where each of $N_{i}$ quarks
have a degenerate mass $m_{v_i}$.
For each valence flavor, the $N_{i}\to 0$ limit has to be taken 
in the end of calculation to complete the partial quenching. 

We parametrize the chiral field in the same way as 
the $\epsilon$ expansion \cite{Gasser:1987ah}, 
by factorizing it into
the zero-momentum mode $U_0$ and non-zero modes $\xi(x)$,
\begin{eqnarray}
\label{eq:param}
U(x)=U_0\exp(i\sqrt{2}\xi(x)/F).
\end{eqnarray}
In our calculation, we perform exact group integration over $U_0$,
while $\xi(x)$ is perturbatively treated always imposing
\begin{eqnarray}
\int d^4 x\; \xi(x) =0,
\end{eqnarray}
to avoid double counting of the zero-mode.

It is known that 
group integration over $U(N_f)$ manifold 
is easier and can be analytically expressed in a simpler form
than the $SU(N_f)$ group case.
For this practical reason,
we consider sectors of fixed topology $Q$, 
which is obtained by the Fourier transform of the partition function,
\begin{eqnarray}
\frac{1}{2\pi}\int_0^{2\pi} d\,\theta\ e^{i\theta Q} \int {\mathcal D} U e^{-{\mathcal L}} .
\end{eqnarray}
We then absorb the $\theta$ integral to
the zero-mode sector:
$e^{i\theta/N_f}U_0 \to U_0 $
and extend our integration
to $U(N_f)$ (or $U(N_f+N)$ in the partially quenched case) group.
The phase factor in the Fourier transform becomes
$e^{i\theta Q} = (\det U_0)^Q.$
The conventional $\theta=0$ vacuum result is obtained by
summing each topological sector with a weight 
given by the partition function, which will be discussed later
in Section~\ref{sec:2pt}.

We give the same counting rule as in the $p$ expansion
for the $\xi$ fields and other parameters,
\begin{eqnarray}
\partial_\mu \sim {\cal O}(p),\;\;\;
\xi(x) \sim {\cal O}(p),\;\;\;
\mathcal{M}\sim {\cal O}(p^2),\;\;\;
T, L \sim {\cal O}(1/p), \label{pcounting}
\end{eqnarray}
in units of the cut-off $4\pi F$.
We assume as usual that the linear sizes of the 4-dimensional volume, $L$ and $T$, 
are much larger than  the inverse QCD scale $\Lambda_{\rm QCD}^{-1}$
so that the effective theory is valid.

According to the counting rule Eq.~(\ref{pcounting}), 
let us expand the Lagrangian 
\begin{eqnarray}
\label{eq:Lchpt}
{\cal L} &=& \frac{1}{2}\mathrm{Tr}(\partial_\mu \xi)^2-\frac{\Sigma}{2}\mathrm{Tr}\left[
\mathcal{M}^\dagger U_0+U_0^\dagger \mathcal{M}\right]
+\frac{1}{2}\sum_i M^2_{ii}[\xi^2]_{ii}
\nonumber\\&&
+\frac{\Sigma}{2F^2}
{\rm Tr}[\mathcal{M}^\dagger (U_0-1)\xi^2+\xi^2(U_0^\dagger-1)\mathcal{M}]+\cdots,
\end{eqnarray}
where $M^2_{ij}=(m_i+m_j)\Sigma/F^2$.
Here we have separated the mass term into three pieces.
The first one (the second term) gives a non-perturbative weight 
in the zero-mode path integration as in the $\epsilon$ expansion 
and the second one (the third term)
has the same form as the conventional mass term (of $\xi$) in the $p$ expansion.

The last term in Eq.~(\ref{eq:Lchpt}) is a mixing term between 
the zero and non-zero modes, which is unfamiliar
either in the $\epsilon$ and $p$ expansions.
In fact, this term plays a crucial role in connecting the
$\epsilon$ and $p$ regimes. We can treat 
this mixing term as a perturbation:
it is not difficult to check 
\begin{eqnarray}
\label{eq:MU-1}
\mathcal{M}(U_0-1) \sim {\cal O}(p^3),
\end{eqnarray}
and, in particular, a Hermitian combination
\begin{eqnarray}
\label{eq:MU-1dag}
\mathcal{M}(U_0+U_0^\dagger-2) \sim {\cal O}(p^4),
\end{eqnarray}
hold in both of the $\epsilon$ and $p$ regimes.
For some specific cases, by a direct group integration, 
one can confirm that these countings are kept even in the intermediate region
where $M_{ij} L\sim 1$ \cite{Damgaard:2008zs}. 
We therefore treat 
Eq.~(\ref{eq:MU-1}) and (\ref{eq:MU-1dag}) 
as the additional counting rules
and treat the last term in Eq.~(\ref{eq:Lchpt})
as an ${\cal O}(p^5)$ contribution.
These additional counting rules 
Eq.~(\ref{eq:MU-1}) and (\ref{eq:MU-1dag}) 
are also supported by the equipartition theorem 
of energy, where the potential energies of 
weekly interacting system are uniformly and
therefore, mass-independently distributed.

In Table~\ref{tab:3exp}, we summarize the difference of 
the three $\epsilon$, $p$, and our new $i$ (=interpolating)
expansions of ChPT.

\begin{table}[tbp]
\centering
\begin{tabular}{|c|c|c|}
\hline\hline
  expansion & parametrization & counting rule\\
\hline
$\epsilon$ expansion & $U(x)=U_0\exp\left(i\frac{\sqrt{2}\xi}{F}\right)$ & 
$U_0\sim {\cal O}(1)$, $\xi\sim {\cal O}(1/L)$, $\mathcal{M}\sim {\cal O}(1/L^4)$\\
\hline
$p$ expansion & $U(x)=\exp\left(i\frac{\sqrt{2}\xi}{F}\right)$ & 
$\xi\sim {\cal O}(1/L)$, $\mathcal{M}\sim {\cal O}(1/L^2)$\\
\hline
$i$ expansion & $U(x)=U_0\exp\left(i\frac{\sqrt{2}\xi}{F}\right)$ & 
$U_0\sim {\cal O}(1)$, $\xi\sim {\cal O}(1/L)$, $\mathcal{M}\sim {\cal O}(1/L^2)$,\\
&&$\mathcal{M}(U_0-1) \sim {\cal O}(1/L^3),\;\;\; \mathcal{M}(U_0+U_0^\dagger-2) \sim {\cal O}(1/L^4)$\\
\hline
\end{tabular}
\caption{Three expansions of ChPT at finite volume.
The counting rules are compared in the units of 
the smallest non-zero momentum $1/L$. 
Our new expansion in this paper is denoted by ``$i$ expansion''.}
\label{tab:3exp}
\end{table}

In the following sections, we calculate
two-point correlation function of 
the peudoscalar operators in three steps.
For the first step (Section~\ref{sec:Leff}), we rewrite the chiral Lagrangian
in terms of non-self-contracting (NSC) vertices of $\xi$ fields.
This corresponds to partly performing one-loop integrals
in the vertices in advance.
By doing this, one can renormalize the
coupling constants and the wave function at NLO
before starting the complicated calculation.
Then the second step for the two-point functions 
(Section \ref{sec:xi}) becomes clearer:
to collect the remaining diagrams,
namely those without self-contractions in vertices,
 which is expressed by
the already renormalized quantities,
and perform $\xi$ integrals.
The third and final step is to perform nonperturbative
$U_0$ integrals.

For the perturbative calculation of $\xi$ fields, 
we use the same Feynman propagator
as in the $p$ expansion
 except that
the zero-momentum mode contribution is removed:
\begin{eqnarray}
\label{eq:xirule}
\langle \xi_{ij}(x)\xi_{kl}(y)\rangle_\xi
&=&\delta_{il}\delta_{jk}\bar{\Delta}(x-y, M^2_{ij})
-\delta_{ij}\delta_{kl}\bar{G}(x-y,M^2_{ii},M^2_{kk}),
\end{eqnarray}
where $\langle \cdots \rangle_{\xi}$ means an integral over $\xi$, 
whose general expression will be discussed later in Sec.~\ref{sec:xi}.
Note that the second term comes from the
constraint ${\rm Tr}\;\xi=0$.
The propagators $\bar{\Delta}$ and $\bar{G}$ are given 
by 
\begin{eqnarray}
\label{eq:Deltabar}
\bar{\Delta}(x,M^2)&=&\frac{1}{V}\sum_{p\neq 0}
\frac{e^{ipx}}{p^2+M^2},\\
\label{eq:Gbar}
\bar{G}(x,M_{ii}^2,M^2_{jj})&=&
\frac{1}{V}\sum_{p\neq 0}\frac{e^{ipx}}{(p^2+M_{ii}^2)(p^2+M_{jj}^2)
\left(\sum^{N_f}_f\frac{1}{p^2+M^2_{ff}}
\right)},
\end{eqnarray}
where the summation is taken over the non-zero 4-momenta
\begin{eqnarray}
p=2\pi(n_t/T, n_x/L, n_y/L, n_z/L),
\end{eqnarray}
with integer $n_\mu$'s except for $p=(0,0,0,0)$.
For the following calculations, where a non-degenerate 
set of valence and sea quark masses is taken,
it is convenient to define a quantity
\begin{eqnarray}
\bar{A}(x,M^2_{ii},M^2_{jj}) &\equiv &
\bar{G}(x,M^2_{ii},M^2_{jj})-\frac{1}{2}
\left[\bar{G}(x,M^2_{ii},M^2_{ii})+\bar{G}(x,M^2_{jj},M^2_{jj})\right].
\end{eqnarray}
Note that both $\bar{A}(x,M^2_{ii},M^2_{jj})$ and its second derivative 
\begin{eqnarray}
\partial_\mu^2 \bar{A}(x,M^2_{ii},M^2_{jj}) &= &
M_{ij}^2\bar{G}(x,M^2_{ii},M^2_{jj})-\frac{1}{2}
\left[M_{ii}^2\bar{G}(x,M^2_{ii},M^2_{ii})
+M_{jj}^2\bar{G}(x,M^2_{jj},M^2_{jj})\right],\nonumber\\
\end{eqnarray}
are UV finite even in the limit $x=0$.
Also, note that both vanish when $M_{jj}^2=M_{ii}^2$.


As a final remark of this section, we note that
the above parametrization Eq.~(\ref{eq:param}) gives rise to a non-trivial
Jacobian in the functional integral measure. 
It is uniquely determined by the left-right invariance of the group integrals.
A perturbative calculation \cite{Hansen:1990un, Bernardoni:2007hi}
has shown that the Jacobian is expressed by
\begin{eqnarray}
\label{eq:measure}
\mathcal{J}(U_0,\xi)&=&\exp\left(-\int d^4x \frac{N_f}{3F^2V}{\rm Tr}\;\xi^2(x)\right),
\end{eqnarray}
to ${\cal O}(p^2)$. 
It plays a role 
just as an additional mass term in our calculation.
\section{Chiral Lagrangian at one-loop}
\label{sec:Leff}

Since our target system is a complicated mixture
of $U_0$ matrix model and perturbative $\xi$-fields,
we first simplify the chiral Lagrangian and 
collect relevant pieces for our computation.
In particular, by introducing non-self-contracting
vertices, we can renormalize (at the one-loop level)
the coupling constants and the $\xi$ fields in advance.

\subsection{Next-to-leading order (NLO) terms}

Without source terms, 
we have eight NLO terms, whose low-energy constants 
are denoted by $L_i$'s ($i=1,\cdots 8$) \cite{Gasser:1983yg}.
In our perturbative expansion 
at ${\cal O}(p^5)$ and ${\cal O}(p^6)$, the terms with 
$L_1, L_2, L_3$ (and the Wess-Zunimo-Witten term 
\cite{Wess:1971yu, Witten:1983tw} as well) 
do not contribute to pseudo-scalar meson masses and decay constants. 
By explicitly expanding $U(x)=U_0 e^{i\sqrt{2}\xi(x)/F}$
in $\xi$, it is sufficient to consider
\begin{eqnarray}
\label{eq:LNLOex}
\mathcal{L}_{NLO}&=&
-\frac{\Sigma}{2}
{\rm Tr}
[\mathcal{M}^{\dagger} U_0
+U_0^\dagger\mathcal{M}]
\times\left(\frac{16 L_6}{F^2}\sum_f M^2_{ff}\right)
\nonumber\\&&
+\sum_{i, j}\left[\frac{1}{2}
\partial_\mu \xi_{ij}
\partial_\mu \xi_{j i}
\right]\times \frac{8}{F^2}\left(
L_4\sum^{N_f}_f M^2_{ff}+L_5M^2_{ij}
\right)
\nonumber\\&&
+\sum_{i, j}\left[\frac{1}{2}\xi_{ij}\xi_{j i}
M^2_{ij}\right]\times\frac{16}{F^2}\left(
L_6\sum_f M^2_{ff}+L_8 M^2_{ij} \right)
+\frac{8L_7}{F^2}\sum^{N_f}_{i,j}M^2_{ii}
M^2_{jj}\xi_{ii}\xi_{jj}
\nonumber\\&&
-4L_8 \sum_i M_{ii}^4\left(\frac{[U_0+U_0^\dagger]_{ii}}{2}-1\right)
-L_8 \sum_{i\neq j}M_{ii}^2M_{jj}^2([U_0]_{ij}[U_0]_{ji}
+[U_0^\dagger]_{ij}[U_0^\dagger]_{ji}).
\end{eqnarray}
Note that we can always omit the constant terms unless source terms are 
inserted (the source insertion is separately discussed below).
It is also important to note in the above expansion 
that the only $L_6$ term has non-trivial $U_0$ dependence at ${\cal O}(p^4)$.

\subsection{Non-self-contracting (NSC) vertices}

For one-loop level calculations, it is convenient to 
rewrite the chiral Lagrangian so that 
quantum corrections are partly included.
This is performed by simply adding and subtracting 
all possible $\xi$-contractions of the $n$ point term
and define the {\it non-self-contracting (NSC) vertex} :
\begin{eqnarray}
\label{eq:nptNSC}
\xi^n(x) &=& [\xi^n(x)]^{NSC} 
+ (\mbox{all possible $\xi$ contractions}),\\
{}[\xi^n(x)]^{NSC} &\equiv&  \xi^n(x)
- (\mbox{all possible $\xi$ contractions}).
\end{eqnarray}
The contracted vertices (second term of Eq.~(\ref{eq:nptNSC}) )
are treated as shifts
of the lower order terms.
These $\xi$ contractions, as they contain the {\it tadpole} diagrams, 
are typically UV divergent. We use the dimensional regularization and 
absorb it into the higher order LECs.
In this way, the coupling renormalization can be 
done in advance, and one can substantially reduce the number of 
remaining one-loop diagrams for an arbitrary correlation function.
Note that $\langle [\xi^n(x)]^{NSC}\rangle_\xi=0$ by definition.

The two-point vertex is the easiest example:
\begin{eqnarray}
\label{eq:2ptNSC}
{}[\xi^2(x)]^{NSC} &=&  \xi^2(x)- \langle \xi^2(x)\rangle_\xi,
\end{eqnarray}
which is applied to the 4-th term of Eq.~(\ref{eq:Lchpt}),
and in this case, the $\xi$ contraction is treated as
a shift of $\Sigma$ in the second term of
Eq.~(\ref{eq:Lchpt}).
Its UV divergence is absorbed into $L_6$.

With the NSC vertices, $L_i$ terms in Eq.~(\ref{eq:LNLOex}), and measure term 
Eq.~(\ref{eq:measure}) together, 
we can express the low-energy effective action as
\begin{eqnarray}
\label{eq:Leff}
\int d^4x\;\mathcal{L} &=&
-\frac{\Sigma_{\rm eff}V}{2}{\rm Tr}
[\mathcal{M}^\dagger U_0+U_0^\dagger \mathcal{M}]
\nonumber\\
&&+\int d^4x\;\frac{1}{2}\sum_{i,j}^{N_f}(Z_\xi^{ij})^2\left(
[\partial_\mu \xi_{ij} \partial_\mu \xi_{ji}]^{NSC}(x)
+\left(M^\prime_{ij}\right)^2
[\xi_{ij}\xi_{ji}]^{NSC}(x)
\right)\nonumber\\
&&
+\mathcal{S}^{(1)}_{I}(U_0,\xi)+\mathcal{S}^{(2)}_{I}(U_0)
\nonumber\\&&+\mathcal{S}_{diag}+\mathcal{S}_{\rm 3pt}+\mathcal{S}_{\rm 4pt},
\end{eqnarray}
where the first two terms are the LO contribution, and
the perturbative interaction terms are given by
\begin{eqnarray}
\label{eq:Smix}
\mathcal{S}^{(1)}_{I}(U_0,\xi)
&\equiv &
\int d^4 x\frac{\Sigma}{2F^2}{\rm Tr}
\left[
\left(\mathcal{M}^\dagger (U_0-1)+(U_0^\dagger-1) \mathcal{M}\right)
[\xi^2(x)]^{NSC}
\right],\\
\mathcal{S}^{(2)}_{I}(U_0)&\equiv &
-\frac{\Sigma V}{2}\sum_i m_i
\left([U_0+U_0^\dagger]_{ii}-2\right) 
\left(-\Delta Z_{ii}^{\Sigma}+\frac{8 L_8}{F^2}M_{ii}^2\right)
\nonumber\\&&
-L_8 V\sum_{i\neq j}M_{ii}^2M_{jj}^2
\left([U_0]_{ij}[U_0]_{ji}
+[U_0^\dagger]_{ij}[U_0^\dagger]_{ji}\right).
\end{eqnarray}
Here we have used notations below,
\begin{eqnarray}
\label{eq:Sigmaeff}
\Sigma_{\rm eff}&\equiv& 
\Sigma \left[
1-\frac{1}{F^2}\left(\sum_f^{N_f}\bar{\Delta}(0,M^2_{ff}/2)
-\bar{G}(0,0,0)\right)
+\frac{16L_6}{F^2}
\sum_f^{N_f}M_{ff}^2\right],\\
\Delta Z^{\Sigma}_{ii} &\equiv& \frac{1}{F^2}
\left(\sum_f^{N_f}(\bar{\Delta}(0,M^2_{if})-\bar{\Delta}(0,M^2_{ff}/2))
-(\bar{G}(0,M_{ii}^2,M_{ii}^2)-\bar{G}(0,0,0))\right),\\
Z_\xi^{ij} &\equiv& 1-\frac{1}{2F^2}
\left[\frac{1}{6}\sum_f^{N_f}(\bar{\Delta}(0,M_{if}^2)
+\bar{\Delta}(0,M_{jf}^2))
+\frac{1}{3}\bar{A}(0,M_{ii}^2,M_{jj}^2)
\right.\nonumber\\&&\left.\hspace{1in}
-8\left(L_4\sum_f^{N_f}M_{ff}^2+L_5M_{ij}^2\right)
\right],
\end{eqnarray}
and $(M^\prime_{ij})^2=(Z_M^{ij}M_{ij})^2+N_f/F^2 V$ with
\begin{eqnarray}
Z_M^{ij} &\equiv& 1+\frac{1}{2F^2}
\left[\bar{G}(0,M_{ii}^2,M_{jj}^2)
-8(L_4-2L_6)\sum_f^{N_f}M_{ff}^2-8(L_5-2L_8)M_{ij}^2
\right].
\end{eqnarray}


In the last line of Eq.~(\ref{eq:Leff}), we have
\begin{eqnarray}
\mathcal{S}_{diag}&\equiv& \int d^4x\;\frac{1}{2F^2}\sum^{N_f}_{i,j} 
\left([\partial_\mu \xi_{ii}\partial_\mu \xi_{jj}]^{NSC}(x)
\frac{\bar{\Delta}(0,M_{ij}^2)}{3}
\right.\nonumber\\&&\left.
-\left(\frac{2}{3}M_{ij}^2\bar{\Delta}(0,M_{ij}^2)
-16L_7M^2_{ii}M^2_{jj}
+\frac{1}{3V}
\right)
[\xi_{ii}\xi_{jj}]^{NSC}(x)\right),\\
\mathcal{S}_{\rm 3pt}&\equiv& 
\int d^4x\;\frac{i\Sigma}{3\sqrt{2}F^3}{\rm Tr}\left[[\xi^3(x)]^{NSC}
\left(\mathcal{M}^\dagger U_0-U_0^\dagger\mathcal{M}\right)\right],\\
\mathcal{S}_{\rm 4pt}&\equiv& \int d^4x\;\left[-\frac{1}{12F^2}
\sum_i^{N_f}M_{ii}^2[\xi^4(x)]^{NSC}_{ii}+\frac{1}{6F^2}{\rm Tr}
[\partial_\mu \xi \xi \partial_\mu \xi \xi
-\xi^2(\partial_\mu \xi)^2]^{NSC}(x)\right],
\end{eqnarray}
but they do not contribute to the calculations in this paper 
where we only consider two-point functions of off-diagonal sources. 
We therefore simply ignore them in the following sections.
We have also ignored trivial constant terms in the above expressions.

\subsection{Pseudoscalar (and scalar) source term}

The pseudoscalar and scalar source terms are
obtained by extending the mass matrix:
\begin{eqnarray}
\label{eq:Psrc}
\mathcal{M}\to \mathcal{M}_J = \mathcal{M}+i\mathcal{J}(x),
\end{eqnarray}
where the pseudoscalar and scalar parts are given by
\begin{eqnarray}
p(x)&=&\frac{1}{2}(\mathcal{J}(x)+\mathcal{J}^\dagger(x)),\\
s(x)&=&\frac{i}{2}(\mathcal{J}(x)-\mathcal{J}^\dagger(x)),
\end{eqnarray}
respectively.

In order to keep a manifest and consistent counting rule,
we treat $\mathcal{M}_J$ in the same way
as the original mass matrix, {\it i.e.},
\begin{eqnarray}
\mathcal{J}(x)\sim {\cal O}(p^2),\;\;\; \mathcal{J}(x)(U_0-1)\sim {\cal O}(p^3),
\;\;\;\mathcal{J}(x)(U_0+U_0^\dagger-2)\sim {\cal O}(p^4).
\end{eqnarray}

Note however  that unlike the original mass matrix,
$\mathcal{J}$-derivative could 
isolate the matrix element of
$(U_0-1)$, which could cause ambiguity 
in the counting rule of correlation functions.
In fact, the leading contribution of the 
pseudoscalar two-point function 
is known to be ${\cal O}(1)$ in the $\epsilon$ expansion while 
it becomes one order higher, ${\cal O}(p^2)$, in the $p$ expansion.
To avoid this problem, we consider 
every $\mathcal{J}_{ij}$-derivative
multiplied by a factor $\sqrt{m_im_j}$ 
\begin{eqnarray}
  \label{eq:MP}
\sqrt{m_im_j}\left[\frac{\delta}{\delta \mathcal{J}(x)}\right]_{ij},
\end{eqnarray}
as a unit block of the calculation. 
This prescription keeps the counting order of 
the operand unchanged even after differentiation.
Note that the unusual square root does not appear 
in the physical results since even numbers of derivatives 
are always required to give a non-zero 
correlation when $i\neq j$.
The pseudoscalar two-point correlation, 
which is our target of this work, is then kept 
always at ${\cal O}(p^6)$ in an unambiguous way 
with arbitrary choice of the quark masses.

Unlike the Lagrangian itself,
we need to introduce an unphysical constant counterterm
with a coefficient $H_2$ \cite{Gasser:1983yg},
\begin{eqnarray}
-H_2  \left(\frac{2\Sigma}{F^2}\right)^2
{\rm Tr} [(\mathcal{M}+i\mathcal{J})^{\dagger}(\mathcal{M}+i\mathcal{J})],
 \end{eqnarray}
to cancel the divergence of the scalar operator at a finite valence quark mass.

Now let us collect terms linear in $\mathcal{J}$
and rewrite it in terms of NSC vertices at ${\cal O}(p^5)$:
\begin{eqnarray}
\label{eq:src}
\mathcal{L}_J &=& 
i\frac{\Sigma_{\rm eff}}{2}
{\rm Tr}[\mathcal{J}^\dagger (x)U_0-U^\dagger_0\mathcal{J}(x)]
\nonumber\\&&
-\frac{\Sigma}{\sqrt{2}F}
\sum_{i,j}^{N_f}\xi_{ij}(x)
[\mathcal{J}^\dagger (x)U_0+U^\dagger_0\mathcal{J}(x)]_{ji}
\times Z_\xi^{ij}Z_F^{ij}(Z_M^{ij})^2
\nonumber\\&&
+i\frac{\Sigma}{2}\sum_{i,j} 
\left(p^\dagger_{ij}(x)[U_0]_{ji}-[U_0^\dagger]_{ij}p_{ji}(x)\right)
\left(-\Delta Z_{ii}^{\Sigma}+\frac{16 L_8}{F^2}M_{ij}^2\right)
\nonumber\\&&
+\Sigma \sum_i^{N_f}s(x)_{ii}
\left(\Delta Z_{ii}^{\Sigma}-\frac{4(2L_8+H_2)M_{ii}^2}{F^2}\right)
\nonumber\\&&
+\frac{\sqrt{2}\Sigma}{3F^3}
\sum_{i,j}^{N_f}p_{ii}(x)\xi_{jj}(x)\bar{\Delta}(0,M_{ij}^2)
-\frac{\sqrt{2}\Sigma}{F}{\rm Tr}[p(x)]
\times\left(\frac{16L_7}{F^2}\sum_f^{N_f}M_{ff}^2\xi_{ff}(x)\right)
\nonumber\\&&
-i\frac{\Sigma}{2F^2}{\rm Tr}[\mathcal{J}^\dagger (x)U_0\xi^2(x)
-\xi^2(x)U^\dagger_0\mathcal{J}(x)]^{NSC},
\end{eqnarray}
where a term  with the cubic NSC vertex $[\xi^3]^{NSC}$ 
is ignored since it never
contributes to the two-point correlation functions.
A new factor $Z_F^{ij}$ is defined by
\begin{eqnarray}
Z_F^{ij}&\equiv& 
1-\frac{1}{2F^2}\left[
\frac{1}{2}\sum_f^{N_f}(\bar{\Delta}(0,M_{if}^2)
+\bar{\Delta}(0,M_{jf}^2))
+\bar{A}(0,M_{ii}^2,M_{jj}^2)
\right.\nonumber\\&&\left.\hspace{1in}
-8\left(L_4\sum_f^{N_f}M_{ff}^2+L_5M_{ij}^2\right)
\right].
\end{eqnarray}

\subsection{Renormalization}

In the above results, $\bar{\Delta}(0,M^2)$ and $\bar{G}(0,M^2_1,M^2_2)$
have the exactly same logarithmic divergences as 
the conventional $p$ expansion since the absence of the zero mode do not affect 
the ultra-violet properties.
In the same way as in \cite{Gasser:1983yg},
we can thus evaluate their divergent parts 
by the dimensional regularization at $D=4-2\epsilon$ (taking $\epsilon \ll 1$):
\begin{eqnarray}
\bar{\Delta}(0,M^2) &=& -\frac{M^2}{16\pi^2}\left(\frac{1}{\epsilon}+1-\gamma+\ln 4\pi\right) +\cdots,\nonumber\\
\bar{G}(0,M_1^2,M_2^2) &=& -\frac{1}{16\pi^2}\left(\frac{M_1^2+M_2^2}{N_f}-
\frac{1}{N_f^2}\sum_f^{N_f}M^2_{ff}\right)
\left(\frac{1}{\epsilon}+1-\gamma+\ln 4\pi\right)
+\cdots,
\end{eqnarray}
where $\gamma=0.57721\cdots$ denotes Euler's constant.
As is the usual case, these divergences can be absorbed into
the renormalization of $L_i$'s and $H_2$ as
\begin{eqnarray}
\label{eq:Lr}
L_i &=& L_i^r(\mu_{sub})-\frac{\gamma_i}{32\pi^2}
\left(\frac{1}{\epsilon}+1-\gamma+\ln 4\pi-\ln \mu^2_{sub}\right),\\
H_2 &=& H_2^r(\mu_{sub})-\frac{\gamma_{H_2}}{32\pi^2}
\left(\frac{1}{\epsilon}+1-\gamma+\ln 4\pi-\ln \mu^2_{sub}\right),
\end{eqnarray}
where $L^r_i(\mu_{sub})$'s and $H_2^r(\mu_{sub})$
denote the renormalized low energy constants
at the subtraction scale $\mu_{sub}$ and
\begin{eqnarray}
\gamma_4 = \frac{1}{8}, \;\;
\gamma_5=\frac{N_f}{8},\;\;
 \gamma_6 = \frac{1}{8}\left(\frac{1}{2}+\frac{1}{N_f^2}\right),\;\;
\gamma_7=0, \;\;
\gamma_8=\frac{\gamma_{H_2}}{2}
=\frac{1}{8}\left(\frac{N_f}{2}-\frac{2}{N_f}\right).
\end{eqnarray}
As a result, $\Sigma_{\rm eff}$, $\Delta Z_{ii}^{\Sigma}$,
$Z^{ij}_F$ and $Z^{ij}_M$ are kept finite, while
$Z^{ij}_\xi$ still diverges 
but it never appears in the physical observables.

After this procedure, one can replace $\bar{\Delta}(0,M^2)$ by,
\begin{eqnarray}
\label{eq:Deltaren}
\bar{\Delta}^r(0,M^2) &=& 
\frac{M^2}{16\pi^2}\ln\frac{M^2}{\mu_{sub}^2}+\bar{g}_1(M^2),
\end{eqnarray}
where $\bar{g}_1$ denotes the finite volume 
contribution of which the
zero-mode part is subtracted.
It is well-known that there are two expressions
for $\bar{g}$: one valid for small $ML\lesssim 1$ 
\cite{Hasenfratz:1989pk}
and the other valid for $ML\gtrsim 1$
\cite{Bernard:2001yj}, 
and their convergence around $ML\sim 1$ is
discussed in detail in Ref.\cite{Damgaard:2008zs}.
Here we just note that on a $L\sim 2$ fm box, these two 
\begin{eqnarray}
\label{eq:g1nume}
\bar{g}_1(M^2) = \left\{
\begin{array}{lc}
\displaystyle \sum_{a
\neq 0}^{|n_i| \leq n^{max}_1} 
\frac{\sqrt{M^2}}{4\pi^2|a|}K_1(\sqrt{M^2}|a|)-\frac{1}{M^2V}& (|M|L>2)\\\\
\displaystyle
-\frac{M^2}{16\pi^2}\ln (M^2 V^{1/2})
-\sum^{n^{max}_2}_{n=1}\frac{\beta_n}{(n-1)!} M^{2(n-1)}V^{(n-2)/2} &(|M|L\leq 2)
\end{array}
\right. ,
\end{eqnarray}
at $n_1^{max}=7$ and $n_2^{max}=300$ 
show a good convergence around 
the threshold $|M|L= 2$.
Here $K_1$ is the modified Bessel function and
the summation is taken over the 4-vector 
$a_\mu = n_\mu L_\mu$ with $L_i = L\; (i=1,2,3)$ and $L_4=T$.
$\beta_i$'s denote the {\it shape coefficients} defined 
in \cite{Hasenfratz:1989pk}.


\section{$\xi$ contractions in the correlator}
\label{sec:xi}

We are now calculating a hybrid system of
a matrix $U_0$ and fields $\xi$
whose partition function (with the source $\mathcal{J}$)
is given by
\begin{eqnarray}
\mathcal{Z}(\mathcal{J}) &=& \int_{U(N_f)} dU_0\;(\det U_0)^{Q}
\int_{SU(N_f)} d\xi \;\exp\left[-\int d^4 x\; 
(\mathcal{L}+\mathcal{L}_{J}) \right],
\end{eqnarray}
where we need to integrate
over both fields.
The integral over $U_0$, in particular, has to be
non-perturbatively performed.
Our strategy of this study is (i) to perturbatively 
calculate $\xi$ fields first,
(ii) then to perform $U_0$ group integrals.

Let us here define two notations
\begin{eqnarray}
\langle O_1(U_0)\rangle_{U_0} 
&\equiv& 
\frac{\displaystyle\int dU_0 \;(\det U_0)^{Q} \;e^{\frac{\Sigma_{\rm eff}V}{2}{\rm Tr}
[\mathcal{M}^\dagger U_0+U_0^\dagger \mathcal{M}]}\;
O_1(U_0)
}
{\displaystyle\int dU_0 \;(\det U_0)^{Q} \;e^{\frac{\Sigma_{\rm eff}V}{2}{\rm Tr}
[\mathcal{M}^\dagger U_0+U_0^\dagger \mathcal{M}]} },\\
\langle O_2(\xi)\rangle_\xi &\equiv& 
\frac{\displaystyle\int d\xi\; e^{-\int d^4x 
\frac{1}{2}\sum_{i,j}(Z_\xi^{ij})^2
[\xi_{ij} (-\partial_\mu^2+M^{\prime 2}_{ij})\xi_{ji}]^{NSC}}
O_2(\xi)}
{\displaystyle\int d\xi\;e^{-\int d^4x 
\frac{1}{2}\sum_{i,j}(Z_\xi^{ij})^2
[\xi_{ij} (-\partial_\mu^2+M^{\prime 2}_{ij})\xi_{ji}]^{NSC}}},
\end{eqnarray}
with which
any correlation function of $U_0$ and $\xi$ (we denote $f(U_0,\xi)$)
can be expressed as
\begin{eqnarray}
\langle f(U_0,\xi)\rangle 
&=& \frac{\left\langle\left\langle f(U_0,\xi)e^{-\mathcal{S}_I^{(1)}(U_0,\;\xi)}
\right\rangle_\xi e^{-\mathcal{S}_I^{(2)}(U_0)}\right\rangle_{U_0}}
{\left\langle\left\langle e^{-\mathcal{S}_I^{(1)}(U_0,\;\xi)}
\right\rangle_\xi e^{-\mathcal{S}_I^{(2)}(U_0)}\right\rangle_{U_0}},
\end{eqnarray}
where the interaction terms $\mathcal{S}^{(i)}_{I}$'s
are treated perturbatively.
Noting $\mathcal{S}^{(1)}_{I}(U_0,\xi) \sim {\cal O}(p)$
and $\mathcal{S}^{(2)}_{I}(U_0) \sim {\cal O}(p^2)$,
the correlation function above at NLO can be
divided into four parts:
\begin{eqnarray}
\langle f(U_0,\xi)\rangle &=& \langle f(U_0,\xi)\rangle^{00}
 +\langle f(U_0,\xi)\rangle^{10}+
\langle f(U_0,\xi)\rangle^{20}+
\langle f(U_0,\xi)\rangle^{01},
\end{eqnarray}
where the superscripts 00,10,20,01 mean
${\cal O}(1)$, ${\cal O}(\mathcal{S}^{(1)}_{I})$,
${\cal O}((\mathcal{S}^{(1)}_{I})^2)$ and
${\cal O}(\mathcal{S}^{(2)}_{I})$, respectively.
Namely, they are defined by
\begin{eqnarray}
\left\langle f(U_0,\xi)\right\rangle^{00} &\equiv&
\left\langle\left\langle f(U_0,\xi)\right\rangle_{\xi}\right\rangle_{U_0},\\
\left\langle f(U_0,\xi)\right\rangle^{10} &\equiv&
\left\langle\left\langle -\mathcal{S}^{(1)}_{I}f(U_0,\xi)\right\rangle_{\xi}\right\rangle_{U_0}
-\left\langle\left\langle -\mathcal{S}^{(1)}_{I}\right\rangle_{\xi}\right\rangle_{U_0}
\left\langle\left\langle f(U_0,\xi)\right\rangle_{\xi}\right\rangle_{U_0},\\
\left\langle f(U_0,\xi)\right\rangle^{20} &\equiv&
\left\langle\left\langle \frac{1}{2}(\mathcal{S}^{(1)}_{I})^2f(U_0,\xi)
\right\rangle_{\xi}\right\rangle_{U_0}
-\left\langle\left\langle \frac{1}{2}(\mathcal{S}^{(1)}_{I})^2\right\rangle_{\xi}\right\rangle_{U_0}
\left\langle\left\langle f(U_0,\xi)\right\rangle_{\xi}\right\rangle_{U_0},\\
\left\langle f(U_0,\xi)\right\rangle^{01} &\equiv&
\left\langle-\mathcal{S}^{(2)}_{I}\left\langle f(U_0,\xi)\right\rangle_{\xi}\right\rangle_{U_0}
-\left\langle-\mathcal{S}^{(2)}_{I}\right\rangle_{U_0}
\left\langle\left\langle f(U_0,\xi)\right\rangle_{\xi}\right\rangle_{U_0}.
\end{eqnarray}
These notations are useful in the following calculation.

In the rest of this section, we calculate the $\langle \cdots \rangle_\xi$ part
using the Feynman rule, Eq.~(\ref{eq:xirule}).

\subsection{Chiral condensate to NLO}

For a warming-up, let us first calculate
the one-point scalar function ({\it i.e.} the chiral 
condensate) to 
the next-to-leading order \cite{Damgaard:2008zs}.
In this case, we consider a pure imaginary
diagonal matrix element of the source
\begin{eqnarray}
[\mathcal{J}]_{ij}=-i\delta_{iv}\delta_{jv}s_{vv}(x).
\end{eqnarray}
In this case, the source term in the Lagrangian is
\begin{eqnarray}
\mathcal{L}_J&=&-\frac{\Sigma_{\rm eff}}{2}s_{vv}(x)
[U_0+U_0^\dagger ]_{vv}
-i\frac{\Sigma}{\sqrt{2}F}s_{vv}(x)
[U_0\xi(x)-\xi(x)U_0^\dagger ]_{vv}
\nonumber\\&&
+\Sigma s_{vv}(x)\left(\Delta Z_{vv}^{\Sigma}-\frac{4(2L_8+H_2)M_{vv}^2}{F^2}\right)
\nonumber\\&&
+\frac{\Sigma}{2F^2}s_{vv}(x)[U_0\xi^2(x)
+\xi^2(x)U^\dagger_0]_{vv}^{NSC}+{\cal O}(p^5),
\end{eqnarray}
where the index $v$ is not summed over.

Now we can calculate the chiral condensate 
of the $v$-th valence quark as follows,
\begin{eqnarray}
m_v \langle \bar{q}_vq_v(x)\rangle &=&
\left.m_v \frac{\delta}{\delta s_{vv}(x)} 
\ln \mathcal{Z}(\mathcal{J})\right|_{s_{vv}=0}
\nonumber\\
&=& m_v\left\langle \frac{\Sigma_{\rm eff}}{2}
[U_0+U_0^\dagger ]_{vv}
-\Sigma 
\left(\Delta Z_{vv}^{\Sigma}-\frac{4(2L_8+H_2)M_{vv}^2}{F^2}\right)\right\rangle^{00}\nonumber\\
&=& m_v\left[\frac{\Sigma_{\rm eff}}{2}
\langle[U_0+U_0^\dagger ]_{vv}\rangle_{U_0}
-\Sigma \left(\Delta Z_{vv}^{\Sigma}-\frac{4(2L_8+H_2)M_{vv}^2}{F^2}\right)
\right],
\end{eqnarray}
where we have used 
$\langle \xi(x)\rangle_\xi =0$ and
$\langle 1\rangle_{U_0}=\langle 1\rangle_\xi=1$.
Note that $\langle \bar{q}_vq_v(x)\rangle^{10}=\langle \bar{q}_vq_v(x)\rangle^{20}=
\langle \bar{q}_vq_v(x)\rangle^{01}=0$ to our order,
which can be easily confirmed by a direct calculation
using the fact $\langle[U_0+U_0^\dagger ]_{vv}\rangle^{01}
=\langle[U_0+U_0^\dagger ]_{vv}-2\rangle^{01}$.
The result is, of course, consistent with Ref.~\cite{Damgaard:2008zs}.

\subsection{Pseudoscalar correlator}

Let us next consider the pseudoscalar source.
In the calculation of meson correlators, 
we take a specific
generator of the chiral group 
which has $v_1 v_2$ and $v_2 v_1$ ($v_1\not= v_2$)
elements only. This choice corresponds to the charged pion 
or general kaon type correlators.
Here $v_i$ denotes the valence quark index 
whose mass is given by $m_{v_i}$.
For simplicity, we omit ``$v$'' in the following:
the indices $v_1$ and $v_2$ are denoted by $1$ and $2$,
and their masses are expressed by $m_1$ and $m_2$, respectively.
Namely, we consider 
\begin{eqnarray}
[\mathcal{J}(x)]_{ij} &=& 
\frac{1}{2}(\delta_{1i}\delta_{j2}+\delta_{2 i}\delta_{1 j})p(x),
\end{eqnarray}
where $p(x)$ is a real classical number.

The pseudoscalar source term in the Lagrangian then becomes
\begin{eqnarray}
\mathcal{L}_J &=& \frac{p(x)}{2}(P^{12}(x)+P^{21}(x)),
\end{eqnarray}
where
\begin{eqnarray}
\label{eq:srcP}
P^{12}(x) &=& i\frac{\Sigma_{\rm eff}}{2}
\left([U_0]_{12}-[U_0^\dagger]_{21}\right)
\left(1-\Delta Z^{\Sigma}_{22}+\frac{16 L_8}{F^2}M_{12}^2\right)
\nonumber\\&&
-\frac{\Sigma}{\sqrt{2}F}\xi_{12}(x)
\left([U_0]_{11}+[U_0^\dagger]_{22}\right)Z_\xi^{12}Z_F^{12}(Z_M^{12})^2
\nonumber\\&&
-\frac{\Sigma}{\sqrt{2}F}\sum_{i\neq 1}
\left([U_0]_{1i}\xi_{i2}(x)+\xi_{2i}(x)[U_0^\dagger]_{i1}\right)\nonumber\\&&
-i\frac{\Sigma}{2F^2}\sum_{i,j}
[\xi^2(x)]_{ij}^{NSC}\left([U_0]_{1i}\delta_{j2}
-\delta_{1i}[U_0^\dagger]_{j2}\right),\\
P^{21}(x) &=& (1\leftrightarrow 2).
\end{eqnarray}

Now we are ready to calculate 
the pseudoscalar-pseudoscalar (PP) correlator,
\begin{eqnarray}
\label{eq:PPdef}
m_1 m_2
\langle P(x)P(0) \rangle &=& 
\left.2 m_1 m_2
\frac{1}{\mathcal{Z}(0)}
\frac{\delta}{\delta p(x)}\frac{\delta}{\delta p(0)}
\mathcal{Z}(\mathcal{J})\right|_{p(x),p(0)=0}
\nonumber\\&=&
m_1 m_2\left[
\frac{1}{2}\langle P^{12}(x)P^{21}(0)\rangle 
+\frac{1}{2}\langle P^{12}(x)P^{12}(0)\rangle
+(1 \leftrightarrow 2)
\right],
\end{eqnarray}
where an overall factor of 2 is introduced to compare with
the corresponding lattice {\it connected} diagram.
Note that the procedure Eq.(\ref{eq:MP}) is performed 
but the factor $m_1 m_2$
will be omitted for simplicity in the following calculation.

Although the number of diagrams we need to calculate
is substantially reduced by using the NSC vertices, 
our calculation is still tedious 
because of the off-diagonal elements of $U_0$ 
in the source term Eq.~(\ref{eq:srcP}),  which 
produces various unusual channels in the correlator.
Every step of calculation is, however, rather straightforward
as in the conventional $p$ expansion, 
except for the use of the $\mathcal{M}(U_0-1)\sim {\cal O}(p^3)$ rule.
We therefore skip the details of the calculation in the main text here.
Instead, we summarize several useful formulas for the computation 
in Appendix~\ref{app:xi}
and present each piece of $\langle P(x)P(0)\rangle^{00}$,
$\langle P(x)P(0)\rangle^{10}$,
$\langle P(x)P(0)\rangle^{20}$ and 
$\langle P(x)P(0)\rangle^{01}$ in Appendix~\ref{app:PP00etc}.
We also use the technique in Appendix~\ref{app:U0pert}.

After relevant one-loop integrals over $\xi$,
the pseudoscalar correlator is given by
\begin{eqnarray}
\label{eq:PPtotal}
\langle P(x)P(0)\rangle &=& \langle P(x)P(0)\rangle^{00}+
\langle P(x)P(0)\rangle^{10}+
\langle P(x)P(0)\rangle^{20}+\langle P(x)P(0)\rangle^{01}
\nonumber\\
&=& -\frac{\Sigma^2}{4}(Z_M^{12}Z_F^{12})^4 \mathcal{C}^{0a}
+\frac{\Sigma^2}{\mu_1+\mu_2}
\left(\frac{\Sigma_{\rm eff}}{\Sigma}-(Z_M^{12}Z_F^{12})^2\right)
\mathcal{C}^{0b}
\nonumber\\&&
+\frac{\Sigma^2}{2}
(\Delta Z_{11}^{\Sigma}-\Delta Z_{22}^{\Sigma})
\mathcal{C}^{0c}
+\left.\frac{\Sigma^2}{2F^2}\right[
(Z_F^{12}(Z_M^{12})^2)^2\mathcal{C}^1
\bar{\Delta}(x,M_{12}^{\prime 2})
\nonumber\\&&
+
\mathcal{C}^2
\left.\left(\frac{\Sigma}{F^2}\partial_{M^2}\right)
\bar{\Delta}(x,M^2)\right|_{M^2=M_{12}^2}
\nonumber\\&&
+
\mathcal{C}^3_{12}
\left(\bar{\Delta}(x,M_{11}^2)-\bar{\Delta}(x,M_{12}^2)\right)
+
\mathcal{C}^3_{21}
\left(\bar{\Delta}(x,M_{22}^2)-\bar{\Delta}(x,M_{12}^2)\right)
\nonumber\\&&
+\sum_{j\neq 1}
\mathcal{C}^4_{1j}
\left(\bar{\Delta}(x,M_{2j}^2)-\bar{\Delta}(x,M_{12}^2)\right)
+\sum_{i\neq 2}
\mathcal{C}^4_{2i}
\left(\bar{\Delta}(x,M_{1i}^2)-\bar{\Delta}(x,M_{12}^2)\right)
\nonumber\\&&
+\mathcal{C}^5 \bar{G}(x,M_{11}^2,M_{22}^2)
\nonumber\\&&
+\mathcal{C}^6_{12} 
\left(\bar{G}(x,M_{11}^2,M_{22}^2)-\bar{G}(x,M_{11}^2,M_{11}^2)\right)
\nonumber\\&&
\left.+\mathcal{C}^6_{21} 
\left(\bar{G}(x,M_{11}^2,M_{22}^2)
-\bar{G}(x,M_{22}^2,M_{22}^2)\right)\right],
\end{eqnarray}
where 
\begin{eqnarray}
\mathcal{C}^{0a} &\equiv&  \left\langle ([U_0]_{12}-[U_0^\dagger]_{21})
([U_0]_{21}-[U_0^\dagger]_{12})
+\frac{1}{2}([U_0]_{12}-[U_0^\dagger]_{21})^2
+\frac{1}{2}([U_0]_{21}-[U_0^\dagger]_{12})^2
\right\rangle_{U_0},
\nonumber\\\\
\mathcal{C}^{0b} &\equiv& \left\langle \frac{[U_0+U_0^\dagger]_{11}}{2}
+\frac{[U_0+U_0^\dagger]_{22}}{2}\right\rangle_{U_0},\\
\mathcal{C}^{0c} &\equiv& \frac{1}{4}
\langle ([U_0]_{12}-[U_0^\dagger]_{21})^2
-([U_0]_{21}-[U_0^\dagger]_{12})^2\rangle_{U_0},\\
\mathcal{C}^1 &\equiv& 
\left\langle
([U_0]_{11}+[U_0^\dagger]_{22})([U_0]_{22}+[U_0^\dagger]_{11})
+\sum^{N_f}_{j\neq 1}[U_0]_{1j}[U_0^\dagger]_{j1}
+\sum^{N_f}_{i\neq 2}[U_0]_{2i}[U_0^\dagger]_{i2}
\right.\nonumber\\&&\left.\hspace{1.2in}
+\frac{1}{2}\left\{([U_0]_{12})^2+([U_0^\dagger]_{21})^2
+([U_0]_{21})^2+([U_0^\dagger]_{12})^2\right\}
\right\rangle_{U_0},\\
\mathcal{C}^2 &\equiv& 
\left\langle 
2([\mathcal{R}]_{11}
+[\mathcal{R}]_{22})
-\sum_{j\neq 1}\frac{
[\mathcal{R}]_{1j}[\mathcal{R}]_{j1}
}{m_j-m_1}
-\sum_{i\neq 2}\frac{
[\mathcal{R}]_{2i}[\mathcal{R}]_{i2}
}{m_i-m_2}\right\rangle_{U_0},\\
\mathcal{C}^3_{ij} &\equiv& 
\frac{1}{2}\langle ([U_0]_{ji})^2+([U_0^\dagger]_{ij})^2\rangle_{U_0}
+\frac{\langle [\mathcal{R}]_{ij}
[U^\dagger_0]_{ij}+[U_0]_{ji}[\mathcal{R}]_{ji}
\rangle_{U_0}}{m_i-m_j}
+\frac{\langle ([\mathcal{R}]_{ij})^2
+([\mathcal{R}]_{ji})^2
\rangle_{U_0}}{2(m_i-m_j)^2},\nonumber\\\\
\mathcal{C}^4_{ij} &\equiv&  
\langle [U_0]_{ij}[U_0^\dagger]_{ji}\rangle_{U_0}
+\frac{\langle 
[\mathcal{R}]_{ji}[U_0]_{ij}+
[\mathcal{R}]_{ij}[U^\dagger_0]_{ji}
\rangle_{U_0}}{m_j-m_i}
+\frac{\langle 
[\mathcal{R}]_{ij}[\mathcal{R}]_{ji}
\rangle_{U_0}}{(m_j-m_i)^2},
\\
\label{eq:C5}
\mathcal{C}^5 &\equiv& 
-\left\langle ([U_0]_{12}+[U_0^\dagger]_{21})
([U_0]_{21}+[U_0^\dagger]_{12})
+\frac{1}{2}([U_0]_{12}+[U_0^\dagger]_{21})^2
+\frac{1}{2}([U_0]_{21}+[U_0^\dagger]_{12})^2
\right\rangle_{U_0},\nonumber\\\\
\mathcal{C}_{ij}^6 &\equiv&  
\frac{1}{2}\langle ([U_0]_{ji}+[U_0^\dagger]_{ij})^2\rangle_{U_0}
+\frac{\langle (
[\mathcal{R}]_{ij}+[\mathcal{R}]_{ji})
([U_0]_{ji}+[U_0^\dagger]_{ij})\rangle_{U_0}}{m_i-m_j}
\nonumber\\&&
+\frac{\langle 2[\mathcal{R}]_{ij}[\mathcal{R}]_{ji}
+([\mathcal{R}]_{ij})^2+
([\mathcal{R}]_{ji})^2\rangle_{U_0}}{2(m_i-m_j)^2},
\end{eqnarray}
where we have used a notation
\begin{eqnarray}
\label{eq:R}
\mathcal{R}&\equiv& \mathcal{M}^\dagger (U_0-1)
+(U_0^\dagger-1)\mathcal{M}.
\end{eqnarray}

One should note that many unusual channels appear 
in Eq.~(\ref{eq:PPtotal}), which is a quite unnatural
situation when just a single particle propagator is expected.
However, one will find in the next section, 
many of them actually disappear,
or many of the coefficients $\mathcal{C}$'s vanish
after integration over $U_0$
\footnote{The readers might wonder if
the integration over $\xi_0$ first 
is then inefficient.
But if we perform $U_0$ integrals first,
we need much more tedious computation over 
$U_0$ than what we will see in Sec.~\ref{sec:U0},
which do not disappear until $\xi$ integration is completed.
We thus believe our order of calculation is easier.
}.

\section{Zero-mode integrals}
\label{sec:U0}

The zero-mode's contribution to 
the so-called {\it graded} partition function of
$n$ bosons and $m$ fermions is analytically known
\cite{Splittorff:2002eb, Fyodorov:2002wq,Splittorff:2003cu},
\begin{equation}
\mathcal{Z}^Q_{n,m}(\{\mu_i\})
=
\frac{\det[\mu_i^{j-1}\mathcal{J}_{Q +j-1}(\mu_i)]_{i,j=1,\cdots n+m}}
{\prod_{j>i=1}^n(\mu_j^2-\mu_i^2)\prod_{j>i=n+1}^{n+m}(\mu_j^2-\mu_i^2)},
\end{equation}
in a fixed topological sector of $Q$ where $\mu_i=m_i\Sigma V$.
Here $\mathcal{J}$'s are defined as
$\mathcal{J}_{Q+j-1}(\mu_i)\equiv (-1)^{j-1} K_{Q+j-1}(\mu_i)$ 
for $i=1,\cdots n$ and 
$\mathcal{J}_{Q+j-1}(\mu_i)\equiv I_{Q+j-1}(\mu_i)$ 
for $i=n+1,\cdots n+m$, 
where $I_\nu$ and $K_\nu$ denote the modified Bessel functions.
Partial quenching is completed by taking the boson masses
to those of the valence fermions at the very end of calculation.

Exact group integrals of various matrix elements over $U_0$
can be calculated by differentiating
the above partition function.
The most basic pieces are
\begin{eqnarray}
\mathcal{S}_v &\equiv& \frac{1}{2}
\left\langle [U_0]_{vv} +[U_0^\dagger]_{vv}\right\rangle_{U_0}
= \lim_{\mu_b \to \mu_v}\frac{\partial}{\partial \mu_v} 
\ln \mathcal{Z}^Q_{1,1+N_f}(\mu_b, \mu_v, \{\mu_{sea}\}),\\
\mathcal{D}_{v} &\equiv& \frac{1}{4}
\left\langle 
\left([U_0]_{v v}+[U_0]_{v v}^\dagger\right)^2 \right\rangle
\nonumber\\
&=&
\frac{1}{\mathcal{Z}^Q_{N_f}(\{\mu_{sea}\})}
\lim_{\mu_{b}\to \mu_{v}}
\frac{\partial^2}{\partial \mu_{v}^2}
\mathcal{Z}^Q_{1,1+N_f}(\mu_{b}, \mu_{v},\{\mu_{sea}\}),\\
\mathcal{D}_{v_1v_2} &\equiv& \frac{1}{4}
\left\langle 
\left([U_0]_{v_1 v_1}+[U_0]_{v_1 v_1}^\dagger\right) 
\left([U_0]_{v_2 v_2}+[U_0]_{v_2 v_2}^\dagger\right) \right\rangle
\nonumber\\
&=&
\frac{1}{\mathcal{Z}^Q_{N_f}(\{\mu_{sea}\})}
\lim_{\mu_{b_1}\to \mu_{v_1}, \mu_{b_2}\to \mu_{v_2}}
\frac{\partial}{\partial \mu_{v_1}}\frac{\partial}{\partial \mu_{v_2}} 
\mathcal{Z}^Q_{2,2+N_f}(\mu_{b_1},\mu_{b_2}, \mu_{v_1},\mu_{v_2},\{\mu_{sea}\}),
\end{eqnarray}
where $\mu_{b_i}$ denotes the bosonic spinor mass and
 $\{\mu_{sea}\}$ indicates a set of sea quark masses 
(normalized by $\Sigma V$). 
Note that $\mathcal{D}_{v_1 v_2}$ and $\mathcal{D}_{v}$ differ
even when $m_{v_1}=m_{v_2}=m_v$.

In Ref.~\cite{Damgaard:2007ep}, 
more non-trivial matrix elements
are calculated in terms of the above $\mathcal{S}$'s and $\mathcal{D}$'s
using the left and right invariance of the group integrals.
Their results are summarized in Appendix~\ref{app:U0}.

Now we can simplify $\mathcal{C}^i$'s in terms of 
$\mathcal{S}$'s and $\mathcal{D}$'s.
Note here that for the leading contribution, namely for
$\mathcal{C}^{0a}$ and $\mathcal{C}^{1}$,
we need to use $\Sigma_{\rm eff}$ instead of $\Sigma$ in the arguments.
We distinguish them by putting a superscript ``${\rm eff}$''
like $\mu^{\rm eff}_i (= m_i \Sigma_{\rm eff}V)$ and
$\mathcal{S}^{\rm eff}_i$.
The results are summarized below.
\begin{eqnarray}
\mathcal{C}^{0a} &=& -\frac{4}{\mu^{\rm eff}_1+\mu^{\rm eff}_2}
\left(\mathcal{S}^{\rm eff}_1+\mathcal{S}^{\rm eff}_2\right),\\
\mathcal{C}^{0b} &=& \mathcal{S}_1+\mathcal{S}_2,\\
\mathcal{C}^{0c} &=& 0,\\
\mathcal{C}^1 &=& 2
\left(1+\mathcal{D}^{\rm eff}_{12}
+\frac{Q^2}{\mu^{\rm eff}_1\mu^{\rm eff}_2}\right),\\
m_1 m_2 \mathcal{C}^2 &=& 
2 m_1 m_2 \left[2m_1(\mathcal{S}_1-1)+
2m_2(\mathcal{S}_2-1)+
\sum_{j\neq 1}\frac{m_1-m_j}{\mu_1+\mu_j}
+ \sum_{i\neq 2}\frac{m_2-m_i}{\mu_2+\mu_i}\right],\\
\mathcal{C}^3_{ij} &=& 0,\\ 
m_{i}\mathcal{C}^4_{ij} 
&=& m_{i}\mathcal{C}^6_{ij} \sim {\cal O}(p^6),\\
\mathcal{C}^5 &=& -\frac{4}{\mu_1-\mu_2}
\left(\mathcal{S}_1-\mathcal{S}_2\right).
\end{eqnarray}
Note that we have used $m_i(\mathcal{S}_i-1)\sim {\cal O}(p^4)$. 

Since the $\mathcal{C}^2$ term contributes only in the $p$ regime,
one can substitute the perturbative expression to $\mathcal{S}_i$ 
\cite{Damgaard:2000di, Damgaard:1999ic}:
\begin{eqnarray}
\mathcal{S}_i &=& 1-
 \sum^{N_f}_{j}\frac{1}{\mu_i+\mu_j}+\frac{Q^2}{2\mu_i^2}+\cdots, 
\end{eqnarray}
and obtain
\begin{eqnarray}
m_1 m_2 \mathcal{C}^2 &=& 
4 m_1m_2\left[-\frac{N_f}{\Sigma V}+
\frac{Q^2}{2\mu_1 \mu_2}(m_1+m_2)+\cdots\right].
\end{eqnarray}
Noting $m_1 m_2 \mathcal{C}^1=4m_1 m_2 + {\cal O}(p^6)$,
the 5th term of Eq.~(\ref{eq:PPtotal}) can be absorbed 
into the 4th term (namely, $\mathcal{C}^1$ term)
by shifting the meson mass as
\begin{eqnarray}
M_{12}^{\prime 2} \to  M_{12}^{\prime 2} 
-\frac{N_f}{F^2V}+\frac{Q^2}{2\mu_1\mu_2}M_{12}^2.
\end{eqnarray}
We recall that an unexpected term $\frac{N_f}{F^2V}$ is found
in the definition of $M_{12}^{\prime}$ but it is now canceled out.

Thus the result can be expressed in a simpler form,
\begin{eqnarray}
\label{eq:PPunintegrated}
\langle P(x)P(0)\rangle_Q &=& 
\Sigma^2(Z_F^{12})^2(Z_M^{12})^2
\frac{\mathcal{S}^{\rm eff}_1+\mathcal{S}^{\rm eff}_2}{\mu_1+\mu_2}
\nonumber\\&&
+\frac{\Sigma^2}{F^2} (Z_F^{12})^2(Z_M^{12})^4
\left(1+\mathcal{D}^{\rm eff}_{12}+\frac{Q^2}{\mu^{\rm eff}_1\mu^{\rm eff}_2}\right)
\bar{\Delta}\left(x,(M^Q_{12})^2\right)
\nonumber\\&&
-\frac{2\Sigma^2}{F^2}\frac{\mathcal{S}_1-\mathcal{S}_2}{\mu_1-\mu_2}
\bar{G}(x, M_{11}^2,M_{22}^2),
\end{eqnarray}
where 
\begin{eqnarray}
\label{eq:MQ}
(M^Q_{ij})^2 &\equiv& M_{ij}^2\left(Z_M^{ij}+\frac{Q^2}{4\mu_i\mu_j}\right)^2.
\end{eqnarray}
\section{Results}
\label{sec:2pt}
\subsection{Pseudoscalar correlator at fixed topology and in
$\theta=0$ vacuum}

Let us take the zero-mode projection, or 
integrate Eq.~(\ref{eq:PPunintegrated}) over three-dimensional space (See eq.(\ref{eq:space_int})), 
\begin{eqnarray}
\label{eq:PPd3x}
\mathcal{PP}(t,m_{1},m_{2})_Q \equiv 
\int d^3 x \langle P(x)P(0)\rangle_Q \hspace{-2in} 
\nonumber\\
&=&
\frac{\Sigma^2(Z_F^{12}Z_M^{12})^4}{F^2(Z_F^{12})^2}\frac{1}{2}
\left(1+\mathcal{D}^{\rm eff}_{12}+\frac{Q^2}{\mu^{\rm eff}_1\mu^{\rm eff}_2}\right)
\frac{\cosh (M_{12}^Q(t-T/2))}{M_{12}^Q\sinh(M_{12}^QT/2)}
\nonumber\\&&
+L^3\frac{\Sigma^2(Z_F^{12}Z_M^{12})^2}{\mu_1+\mu_2}
\left[\mathcal{S}^{\rm eff}_1+\mathcal{S}^{\rm eff}_2
-\left(1+\mathcal{D}^{\rm eff}_{12}
+\frac{Q^2}{\mu^{\rm eff}_1\mu^{\rm eff}_2}\right)
\left/\left(1+\frac{Q^2}{2\mu_1\mu_2}\right.\right)
\right]
\nonumber\\&&
-\frac{2\Sigma^2}{F^2}\frac{\mathcal{S}_1-\mathcal{S}_2}{\mu_1-\mu_2}r_{12}(t),
\end{eqnarray}
which is more useful to compare with lattice QCD results,
where 
\begin{eqnarray}
r_{ij}(t) &\equiv& \int d^3 x \;\bar{G}(x,M_{ii}^2,M_{jj}^2).
\end{eqnarray}
This is our main result in this paper valid 
for an arbitrary number of non-degenerate flavors.

It is also important to consider 
the correlator in the $\theta=0$ vacuum,
\begin{eqnarray}
\mathcal{PP}(t,m_{1},m_{2})_{\theta=0} \equiv 
\int d^3 x \langle P(x)P(0)\rangle_{\theta=0}\hspace{-3in} 
\nonumber\\
&=&
\frac{\Sigma^2(Z_F^{12}Z_M^{12})^4}{F^2(Z_F^{12})^2}\frac{1}{2}
\left(1+(\mathcal{D}^{\rm eff}_{12})_{\theta=0}
+\frac{(Q^2)_{\theta=0}}{\mu^{\rm eff}_1\mu^{\rm eff}_2}\right)
\frac{\cosh (M_{12}^{\theta=0}(t-T/2))}{M_{12}^{\theta=0}\sinh(M_{12}^{\theta=0}T/2)}
\nonumber\\&&
+L_3\frac{\Sigma^2(Z_F^{12}Z_M^{12})^2}{\mu_1+\mu_2}
\left[(\mathcal{S}^{\rm eff}_1)_{\theta=0}
+(\mathcal{S}^{\rm eff}_2)_{\theta=0}
-\left(1+(\mathcal{D}^{\rm eff}_{12})_{\theta=0}
+\frac{(Q^2)_{\theta=0}}{\mu^{\rm eff}_1\mu^{\rm eff}_2}\right)
\left/\left(1+\frac{(Q^2)_{\theta=0}}{2\mu_1\mu_2}\right)\right.
\right]
\nonumber\\&&
-\frac{2\Sigma^2}{F^2}\frac{(\mathcal{S}_1)_{\theta=0}
-(\mathcal{S}_2)_{\theta=0}}{\mu_1-\mu_2}r_{12}(t),
\end{eqnarray}
where
$(M^{\theta=0}_{ij})^2 
\equiv M_{ij}^2\left(Z_M^{ij}
+\frac{(Q^2)_{\theta=0}}{4\mu_i\mu_j}\right)^2$.
The summation over topology,
\begin{eqnarray}
\label{eq:sumQ}
(\mathcal{O})_{\theta=0} &\equiv& 
\frac{\sum_Q \mathcal{O}(Q) \mathcal{Z}^Q_{0,N_f}(\{\mu^{\rm eff}_i\})}
{\sum_Q \mathcal{Z}^Q_{0,N_f}(\{\mu^{\rm eff}_i\})},
\end{eqnarray}
can be, at least, numerically performed using the 
analytic expression for 
$\mathcal{Z}^Q_{0,N_f}(\{\mu^{\rm eff}_i\})$, which is finite.
For small $N_f$ cases, simple analytic forms 
are also known \cite{Lenaghan:2001ur}.
Note in the $p$ regime,
that we can easily calculate
$(Q^2)_{\theta=0}=\bar{\mu}\equiv \bar{m}\Sigma V = \chi_t V$
where $\bar{m}=1/\sum_f (1/m_f)$ and $\chi_t$ denotes 
the topological susceptibility
\footnote{In the $p$ regime, 
the LO calculation of $\chi_t$ is enough in this work.
See Refs.~\cite{Aoki:2009mx, Mao:2009sy} for the NLO correction.
}.

As seen above, we find 
a constant contribution in the pseudoscalar correlator 
in addition to the conventional $\cosh$ function of time $t$.
This constant term is indispensable for
 keeping the result IR finite and giving a smooth interpolation
between the $\epsilon$ and $p$ regime limits.

\subsection{Check in the $p$ regime and $\epsilon$ regime limits}

Let us confirm whether our above formulas recover
the conventional $p$ expansion results 
when both of $m_1$, $m_2$ are large (or $m_1, m_2\gg 1/\Sigma V$).
In that limit, we can use 
(see Appendix~\ref{app:U0} and Refs.\cite{Damgaard:2000di, Damgaard:1999ic})
\begin{eqnarray}
\frac{1}{\mu_1+\mu_2}&\sim& {\cal O}(p^2),\\
\mathcal{S}_i &\sim& 1-\sum_f \frac{1}{\mu_i+\mu_f}+\frac{Q^2}{2\mu_i^2}
+{\cal O}(p^4),\\
\mathcal{D}_{12}&\sim& \mathcal{S}_1\mathcal{S}_2
\sim 1 -\sum_f \frac{1}{\mu_1+\mu_f}+\frac{Q^2}{2\mu_1^2}
-\sum_f \frac{1}{\mu_2+\mu_f}+\frac{Q^2}{2\mu_2^2}+{\cal O}(p^4),\\
(Q^2)_{\theta =0} &\sim & \bar{\mu}.
\end{eqnarray}

Here one should remember that in the conventional $p$ expansion,
$Z$ factors are expressed not by $\bar{\Delta}(0,M^2)$ but
by $\Delta(0,M^2)=\bar{\Delta}(0,M^2)+1/M^2V$. 
To take this into account, it is useful to redefine
the $Z$ factors,
\begin{eqnarray}
[Z_M^{12}]_p &\equiv& Z_M^{12}\left[1+\frac{\bar{\mu}}{4\mu_1\mu_2}\right],\\
{}[Z_F^{12}]_p &\equiv& Z_F^{12}\left[1
-\frac{1}{4}\sum_f \left(\frac{1}{\mu_1+\mu_f}
+\frac{1}{\mu_2+\mu_f}\right)
+\frac{1}{8}\left(\frac{1}{\mu_1}-\frac{1}{\mu_2}\right)^2\bar{\mu}\right].
\end{eqnarray}
The well-known result in the $p$ expansion is then precisely recovered,
\begin{eqnarray}
\label{eq:PPptheta}
\mathcal{PP}(t,m_{1},m_{2})_{\theta=0} 
&=&
\frac{\Sigma^2([Z_F^{12}]_p[Z_M^{12}]_p)^4}{F^2([Z_F^{12}]_p)^2}
\frac{\cosh (M_{12}^{\theta=0}(t-T/2))}
{M_{12}^{\theta=0}\sinh(M_{12}^{\theta=0}T/2)},
\end{eqnarray}
where $M_{12}^{\theta=0} = M_{12}[Z_M^{12}]_p$.
Note that the constant term and $r_{12}(t)$ term 
rapidly vanish as $m_1$ or $m_2$ grows.
We also confirm that our result at fixed topology 
agrees with the one in the $p$ expansion \cite{Aoki:2009mx}.\\


Next let us consider the $\epsilon$ regime limit,
where both of the valence masses are near the chiral limit,
$m_1\sim m_2\sim 1/\Sigma V$.
In this case, one can expand the hyperbolic cosine term
in the meson mass as
\begin{eqnarray}
\frac{\cosh (M(t-T/2))}{M\sinh(MT/2)}
&=& \frac{2}{M^2 T}+2Th_1(t/T)+{\cal O}(M^2),
\end{eqnarray}
where 
\begin{eqnarray}
h_1(t/T)&\equiv& 
\frac{1}{2}\left(\frac{t}{T}-\frac{1}{2}\right)^2
-\frac{1}{24},
\end{eqnarray}
and obtains
\begin{eqnarray}
\label{eq:PPepsilon}
\mathcal{PP}(t,m_{1},m_{2})_Q 
&=&
\frac{\Sigma_{\rm eff}^2}{F^2(Z_F^{12})^2}
\left(1+\mathcal{D}^{\rm eff}_{12}+\frac{Q^2}{\mu^{\rm eff}_1\mu^{\rm eff}_2}\right)
Th_1(t/T)
\nonumber\\&&
+L^3\Sigma_{\rm eff}^2
\frac{\mathcal{S}^{\rm eff}_1+\mathcal{S}^{\rm eff}_2}
{\mu^{\rm eff}_1+\mu^{\rm eff}_2}
-\frac{2\Sigma^2}{F^2}\frac{\mathcal{S}_1-\mathcal{S}_2}{\mu_1-\mu_2}r_{12}(t),
\end{eqnarray}
which is consistent with the result in the $\epsilon$ expansion 
(Ref.~\cite{Bernardoni:2008ei}).
Note that we have used
$\Sigma(Z_M^{12}Z_F^{12})^2 = 
\Sigma_{\rm eff}+{\cal O}(M_{11}^2)+{\cal O}(M_{22}^2)$.

\subsection{When $m_2$ is large}
\label{sec:mixedregime}

One of our main interests in this work is to consider 
when one valence quark is always large,
in the $p$ regime: 
$m_2 \Sigma V \sim {\cal O}(1/p^2)$.
Namely, we consider the chiral limit of the 
kaon-type correlators in a finite box.

In this case, we can perturbatively treat (see Appendix~\ref{app:U0})
\begin{eqnarray}
\frac{1}{\mu_1+\mu_2}&\sim& \frac{1}{\mu_2}\sim {\cal O}(p^2),\\
\mathcal{S}_2 &\sim& 1-\sum_f \frac{1}{\mu_2+\mu_f}+\frac{Q^2}{2\mu_2^2}
+{\cal O}(p^4),\\
\mathcal{D}_{12}&\sim& \mathcal{S}_1\left(
1 -\sum_f \frac{1}{\mu_2+\mu_f}+\frac{Q^2}{2\mu_2^2}\right),
\end{eqnarray}
and the correlator in that limit is
\begin{eqnarray}
\label{eq:kaontype}
\mathcal{PP}(t,m_{1},m_{2})_Q 
&=&
\frac{\Sigma^2(Z_F^{12}Z_M^{12})^4}{F^2(Z_F^{12})^2}
\left[1+\mathcal{S}^{\rm eff}_{1}\left(1-\sum_f \frac{1}{\mu_2+\mu_f}
+\frac{Q^2}{2\mu_2^2}\right)
+\frac{Q^2}{\mu_1\mu_2}\right]
\nonumber\\&&
\times
\frac{\cosh (M_{12}^Q(t-T/2))}{2M_{12}^Q\sinh(M_{12}^QT/2)}.
\end{eqnarray}
The result in the $\theta =0$ case is obtained by
replacing $Q^2$ with $(Q^2)_{\theta =0}$ and  $\mathcal{S}_v^{\rm eff}$ with $(\mathcal{S}_v^{\rm eff})_{\theta =0}$.

One can see that the overall factor (and therefore
the calculation of the decay constant $F_K$) 
still has a large finite volume correction 
from the zero-mode integration, 
while the meson mass ($M_{12}^Q$ here) 
has a rather small perturbative correction.

\subsection{Origin of the $\bar{G}(x,M_{11}^2,M_{22}^2)$ term}

The third term in Eq.~(\ref{eq:PPd3x}) 
becomes significant
only when both of $m_1$ and $m_2$ are in the $\epsilon$ regime.
Here we consider the origin of that term.

Although non-perturbative integration of the zero-mode 
is supposed to be the most reliable way of 
calculating the finite size effects near the chiral limit, 
it obscures the physical meaning as 
propagation of the pions.
Let us here go back to a perturbative picture in 
the definition of Eq.~(\ref{eq:C5})
and express the corresponding correlation function using
Appendix~\ref{app:U0pert}
and putting labels ``$(x)$'' and ``$(y)$'' to explicitly 
show where the original operators are located.
For example, the first term of Eq.~(\ref{eq:C5}) is expressed by
\begin{eqnarray}
\left\langle\left([U_0(x)]_{12}+[U_0^\dagger(x)]_{21}\right)
\left([U_0(y)]_{21}+[U_0^\dagger(y)]_{12}\right)\right\rangle_{U_0}
\hspace{-2in}&&\nonumber\\
&=& -\frac{2}{F^2}\left(
\langle[\xi_0(x)]_{12}[\xi_0(y)]_{21}\rangle_{U_0}
+\langle[\xi_0(x)]_{21}[\xi_0(y)]_{12}\rangle_{U_0}
\right)\nonumber\\
&&+\frac{1}{F^4}\left
\langle([\xi^2_0(x)]_{12}+[\xi^2_0(x)]_{21})
([\xi^2_0(y)]_{21}+[\xi^2_0(y)]_{12})\right\rangle_{U_0}
\nonumber\\&&
+\frac{2}{3F^4}\left(
\langle([\xi_0(x)]_{12}-[\xi_0(x)]_{21})
([\xi^3_0(y)]_{21}-[\xi^3_0(y)]_{12})\rangle_{U_0}
\right.\nonumber\\&&\left.
+\langle([\xi^3_0(x)]_{12}-[\xi^3_0(x)]_{21})
([\xi_0(y)]_{21}-[\xi_0(y)]_{12})\rangle_{U_0}
\right)+\cdots.
\end{eqnarray}
With this perturbative picture of the zero-mode,
the $\mathcal{C}^5$ term can be expressed as
\begin{eqnarray}
\mathcal{C}^5\bar{G}(x-y,M_{11}^2,M_{22}^2)
&=& \frac{1}{F^4}
\left\langle([\xi^2_0(x)]_{12}+[\xi^2_0(x)]_{21})
([\xi^2_0(y)]_{21}+[\xi^2_0(y)]_{12})\right\rangle_{U_0}
\nonumber\\&&
\;\;\;\;\;\;\times \left\langle ([\xi(x)]_{11}[\xi(y)]_{22}
+[\xi(x)]_{22}[\xi(y)]_{11})\right\rangle_\xi+\cdots.
\end{eqnarray}
It is then obvious that this term is originally
a three-pion-state propagator which is suppressed 
in the ordinary $p$ regime. 
As the system enters the $\epsilon$ regime, however, two of their zero-mode's 
contributions are non-perturbatively enhanced
and it becomes an NLO contribution.

\section{Useful examples}
\label {sec:example}

In this section we present two specific examples in the
$N_f=2$ (with degenerate up and down quarks )
and $2+1$ (with up, down and strange quarks) 
theories, which 
are useful to analyze lattice QCD results
simulated in finite volumes.
In the formulas below, we denote the sea quark
masses by $m_{u}=m_d=m$ and  $m_s$
($\mu = m\Sigma V$ and $\mu_s = m_s\Sigma V$).

We consider two-types of the pseudoscalar correlators:
the pion-type
correlator whose two valence masses are
degenerate, $m_1=m_2=m_v$ ($\mu_v=m_v\Sigma V$), 
and the kaon-type correlator for which we take $m_2$ always to be
in the $p$ regime 
(see the general formula Eq.~(\ref{eq:kaontype})). 

\subsection{Simplified $\bar{G}(x,M^2,M^2)$}

For small $N_f$, we can simplify the $\bar{G}(x,M_{11}^2, M_{22}^2)$
 (or $r_{12}(t)$) term.
Since it contributes only when both of $m_1$ and $m_2$
are in the $\epsilon$ regime, it is sufficient to
consider the pion-type correlator case with $m_1=m_2=m_v$.
The result was already presented in Ref.~\cite{Aoki:2009mx},  
except for the presence of  the zero-mode part :
$G(x,M_{11}^2, M_{22}^2)=
\bar{G}(x,M_{11}^2, M_{22}^2)+1/(VM_{11}^2M_{22}^2(\sum_f 1/M_{ff}^2))$,
which does not affect the coefficient of each term.
Here we just present the results for the $N_f=2$ and 2+1 cases,
\begin{eqnarray}
\bar{G}(x,M_{vv}^2, M_{vv}^2) &=&
\left\{\begin{array}{ll}
{\displaystyle \frac{1}{2}}\left[\bar{\Delta}(x,M_{vv}^2)
+(M_{vv}^2-M_{\pi}^2)\partial_{M_{vv}^2}\bar{\Delta}(x,M_{vv}^2)\right] 
\;\;\;\;\;\; (N_f=2),
\\\\
{\displaystyle \frac{1}{3}}\left[{\displaystyle -\frac{1}{2}
\frac{(M_\pi^2-M_\eta^2)^2}{(M_{vv}^2-M_\eta^2)^2}}
\bar{\Delta}(x,M_\eta^2)
+\left(1+{\displaystyle \frac{1}{2}\frac{(M_\pi^2-M_\eta^2)^2}
{(M_{vv}^2-M_\eta^2)^2}}\right)
\bar{\Delta}(x,M_{vv}^2)
\right.& \\\left.
+{\displaystyle \frac{(M_{vv}^2-M_{\pi}^2)\left\{3(M_{vv}^2-M_\eta^2)
-(M_{vv}^2-M_\pi^2)\right\}}
{2(M_{vv}^2-M_\eta^2)}}
\partial_{M_{vv}^2}\bar{\Delta}(x,M_{vv}^2)\right] 
\\
\hspace{3in}(N_f=2+1),
\end{array}
\right.\nonumber\\
\end{eqnarray}
where $M_{vv}^2=2m_v \Sigma/F^2$, $M_\pi^2=2m\Sigma/F^2$,
and $M_\eta^2=(2m+4m_s)\Sigma/3F^2$.
Noting that $\bar{\Delta}(x,M_\eta^2)$ rapidly converges to
$-1/M_\eta^2V$ for large $|x|$ and 
remembering that the corresponding term contributes only when
$M_{vv}^2 \sim {\cal O}(p^4)$, it is sufficient to consider 
(See also Eq.(\ref{eq:space_int}).)
\begin{eqnarray}
\label{eq:r2and2+1}
r_{vv}(t) 
&\simeq& \left\{
\begin{array}{ll}
{\displaystyle
\frac{1}{2}\left[
\frac{\cosh(M^Q_{vv}(t-T/2))}{2M^Q_{vv} \sinh(M^Q_{vv}T/2)}
-\frac{1}{(M^Q_{vv})^2T}\right]}&(N_f=2),\\\\
{\displaystyle
\frac{1}{2}\left[
\frac{\cosh(M^Q_{vv}(t-T/2))}{2M^Q_{vv} \sinh(M^Q_{vv}T/2)}
-\frac{1}{(M^Q_{vv})^2T}\right]
+\frac{1}{6M_\eta^2T}}&(N_f=2+1).
\end{array}\right.
\end{eqnarray}
Here we have used an additional assumption that the valence 
pion mass is not taken very differently from the physical pion mass and
the ${\cal O}(M_{vv}^2-M_\pi^2)$ contribution is ignored.
The only exceptional case : $M_\pi^2 \gg M_{vv}^2$ will be
discussed later. Note that we have replaced 
the tree-level mass $M_{vv}$ by the NLO mass 
$M_{vv}^Q$ for later convenience (the difference is NNLO.).

\subsection{$N_f=2$ and $2+1$ flavor results}

Using Eq.~(\ref{eq:r2and2+1}), the pion-type correlator can be 
expressed in a compact form,
\begin{eqnarray}
\label{eq:pion}
\mathcal{\pi}(t,m_{v})_Q\equiv \mathcal{PP}(t,m_{v},m_{v})_Q &=&
C_{PP}^Q  \frac{\cosh (M_{vv}^Q(t-T/2))}{M_{vv}^Q\sinh(M_{vv}^QT/2)}
+D_{PP}^Q,
\end{eqnarray}
where the valence pion mass is given by
\begin{eqnarray}
M_{vv}^Q &=& M_{vv} Z_M^{vv} \left(1+\frac{Q^2}{4\mu_v^2}\right),
\end{eqnarray}
and 
\begin{eqnarray}
C_{PP}^Q &=& \frac{\Sigma^2}{F^2}\frac{(Z_M^{vv}Z_F^{vv})^4}{(Z_F^{vv})^2}
\frac{1}{2}\left(1+\mathcal{D}_{vv}^{\rm eff}
+\frac{Q^2}{(\mu_{v}^{\rm eff})^2}
-\frac{\partial \mathcal{S}_v^{\rm eff}}{\partial \mu^{\rm eff}_v}\right),\\
D_{PP}^Q &=& \left\{
\begin{array}{ll}
{\displaystyle 
L^3\frac{\Sigma_{\rm eff}^2}{2\mu^{\rm eff}_v}
\left[2\mathcal{S}_v^{\rm eff}
-\left(1+\mathcal{D}_{vv}^{\rm eff}
+\frac{Q^2}{(\mu_{v}^{\rm eff})^2}
-\frac{\partial \mathcal{S}_v^{\rm eff}}{\partial \mu^{\rm eff}_v}\right)
\left/\left(1+\frac{Q^2}{2\mu_{v}^2}\right)\right.
\right]} & (N_f=2),
\\\\
{\displaystyle 
L^3\frac{\Sigma_{\rm eff}^2}{2\mu^{\rm eff}_v}
\left[2\mathcal{S}_v^{\rm eff}
-\left(1+\mathcal{D}_{vv}^{\rm eff}
+\frac{Q^2}{(\mu_{v}^{\rm eff})^2}
-\frac{\partial \mathcal{S}_v^{\rm eff}}{\partial \mu^{\rm eff}_v}\right)
\left/\left(1+\frac{Q^2}{2\mu_{v}^2}\right)\right.
\right.}\\\left.
\displaystyle \hspace{1in}
-\frac{\mu^{\rm eff}_v}{\mu^{\rm eff}+2\mu^{\rm eff}_s}
\left(\frac{\partial \mathcal{S}_v^{\rm eff}}{\partial \mu^{\rm eff}_v}\right)
\right] & (N_f=2+1),\\
\end{array}\right.
\end{eqnarray}
Here we have used $\Sigma_{\rm eff}=\Sigma (Z_M^{vv}Z_F^{vv})^2+{\cal O}(m_v)$ and 
\begin{equation}
\lim_{m_1\to m_2=m_v}
\frac{\mathcal{S}_1-\mathcal{S}_2}{\mu_1-\mu_2} = 
\frac{\partial \mathcal{S}_v}{\partial \mu_v}.
\end{equation}

It is also possible to simplify the kaon-type correlator 
(here we choose the second valence mass to be the physical 
strange quark mass: $m_2=m_s$ in the 2+1-flavor theory) as 
\begin{eqnarray}
\label{eq:kaon}
\mathcal{K}(t,m_v)_Q\equiv \mathcal{PP}(t,m_v,m_s)_Q &=&
E_{PP}^Q  \frac{\cosh (M_{vs}^Q(t-T/2))}{M_{vs}^Q\sinh(M_{vs}^QT/2)},
\end{eqnarray}
where the valence kaon mass is given by
\begin{eqnarray}
M_{vs}^Q &=& M_{vs} Z_M^{vs} \left(1+\frac{Q^2}{4\mu_v\mu_s}\right),
\;\;\;\;\;\; M_{vs}=\sqrt{(m_v+m_s)\Sigma/F^2},
\end{eqnarray}
and 
\begin{eqnarray}
E_{PP}^Q &=& \frac{\Sigma^2}{F^2}\frac{(Z_M^{vs}Z_F^{vs})^4}{(Z_F^{vs})^2}
\frac{1}{2}\left(1+\mathcal{S}_{v}^{\rm eff}
\left(1-\frac{2}{\mu_s+\mu}-\frac{1}{2\mu_s}+\frac{Q^2}{2\mu_s^2}\right)
+\frac{Q^2}{\mu_{v}\mu_{s}}
\right).
\end{eqnarray}

The result in the $\theta=0$ vacuum is obtained
by simply replacing $Q^2$ with $(Q^2)_{\theta =0}$, $\mathcal{S}_v^{\rm eff}$ with
$(\mathcal{S}_v^{\rm eff})_{\theta =0}$
and $\mathcal{D}_{vv}^{\rm eff}$ with $(\mathcal{D}_{vv}^{\rm eff})_{\theta =0}$
in the above formulas.

Using a notation for the {\it renormalized} logarithmic term
which is given in Eq.~(\ref{eq:Deltaren}),
the explicit forms of $Z$ factors \cite{Sharpe:2000bc, Bijnens:2006jv}, 
$\Sigma_{\rm eff}/\Sigma$,
$\mathcal{S}_v$ and $\mathcal{D}_{vv}$ (see Appendix \ref{app:U0})
are given by 
\begin{itemize}
\item {$N_f=2$ case :}\\
\end{itemize}
\begin{eqnarray}
Z_M^{vv}&=&1+\frac{1}{2F^2}\left[
\frac{1}{2}\bar{\Delta}^r(0,M_{vv}^2)+\frac{1}{2}(M_{vv}^2-M_\pi^2)
\partial_{M_{vv}^2}\bar{\Delta}^r(0,M_{vv}^2)
\right.\nonumber\\&&\left.\hspace{1in}
-16(L_4^r-2L_6^r)M_\pi^2-8(L_5^r-2L_8^r)M_{vv}^2
\right],\\
Z_F^{vv}&=&1-\frac{1}{2F^2}\left[
2\bar{\Delta}^r(0,(M_{vv}^2+M_\pi^2)/2)
-8(2L_4^rM_\pi^2+L_5^rM_{vv}^2)
\right],\\
\frac{\Sigma_{\rm eff}}{\Sigma} &=& 
1-\frac{1}{F^2}\left[
2\bar{\Delta}^r(0,M_\pi^2/2)
-\frac{1}{2}\left\{-\frac{\beta_1}{\sqrt{V}}-M_\pi^2
\left(-\frac{1}{16\pi^2}\ln V^{1/2}\mu_{sub}^2-\beta_2\right)
\right\}
-32L_6^rM_\pi^2
\right],\nonumber\\\\
\label{eq:Sigma-2flavor}
\mathcal{S}_v &=& 
-\frac{1}{(\mu^{2}-\mu_v^2)^2}
\times \frac{\det \left(
\begin{array}{cccc}
\partial_{\mu_v}K_Q(\mu_{v}) & I_Q(\mu_v) & I_Q(\mu) & 
\mu^{-1}I_{Q-1}(\mu) \\
-\partial_{\mu_v}(\mu_{v} K_{Q+1}(\mu_{v})) 
& \mu_v I_{Q+1}(\mu_v) & \mu I_{Q+1}(\mu) & I_{Q}(\mu)\\
\partial_{\mu_v}(\mu_{v}^2 K_{Q+2}(\mu_{v})) & \mu_v^2I_{Q+2}(\mu_v) & 
\mu^2I_{Q+2}(\mu) 
& \mu I_{Q+1}(\mu) \\
-\partial_{\mu_v}(\mu_{v}^3 K_{Q+3}(\mu_{v})) & \mu_v^3I_{Q+3}(\mu_v) & 
\mu^3I_{Q+3}(\mu) 
& \mu^2 I_{Q+2}(\mu) \\
\end{array}
\right)}{\det \left(
\begin{array}{cc}
I_Q(\mu) & \mu^{-1}I_{Q-1}(\mu) \\
\mu I_{Q+1}(\mu) & I_{Q}(\mu)\\
\end{array}
\right)},\nonumber\\\\
\label{eq:Dvv 2-flavor}
\mathcal{D}_{vv} &=& 
- \frac{1}{(\mu^{2}-\mu_v^2)^2}
\nonumber\\&&
\times \frac{\det \left(
\begin{array}{cccc}
\partial_{\mu_v}K_Q(\mu_{v}) 
& \partial_{\mu_v}I_Q(\mu_v) & I_Q(\mu) & 
\mu^{-1}I_{Q-1}(\mu) \\
-\partial_{\mu_v}(\mu_{v} K_{Q+1}(\mu_{v})) 
& \partial_{\mu_v}(\mu_v I_{Q+1}(\mu_v)) & \mu I_{Q+1}(\mu) & I_{Q}(\mu)\\
\partial_{\mu_v}(\mu_{v}^2 K_{Q+2}(\mu_{v})) & 
\partial_{\mu_v}(\mu_v^2I_{Q+2}(\mu_v)) & 
\mu^2I_{Q+2}(\mu) 
& \mu I_{Q+1}(\mu) \\
-\partial_{\mu_v}(\mu_{v}^3 K_{Q+3}(\mu_{v})) & 
\partial_{\mu_v}(\mu_v^3I_{Q+3}(\mu_v)) & 
\mu^3I_{Q+3}(\mu) 
& \mu^2 I_{Q+2}(\mu) \\
\end{array}
\right)}{\det \left(
\begin{array}{cc}
I_Q(\mu) & \mu^{-1}I_{Q-1}(\mu) \\
\mu I_{Q+1}(\mu) & I_{Q}(\mu)\\
\end{array}
\right)}
\nonumber\\&&
+ \frac{4\mu_v}{\mu^2-\mu_v^2}\mathcal{S}_v.
\end{eqnarray}
\\
\begin{itemize}
\item {$N_f=2+1$ case :}\\
\end{itemize}
\begin{eqnarray}
Z_M^{vv}&=&1+\frac{1}{2F^2}\left[
-\frac{1}{6}\frac{(M_\pi^2-M_\eta^2)^2}{(M_{vv}^2-M_\eta^2)^2}
\bar{\Delta}^r(0,M_\eta^2)+\frac{1}{3}\left(1+
\frac{1}{2}\frac{(M_\pi^2-M_\eta^2)^2}{(M_{vv}^2-M_\eta^2)^2}\right)
\bar{\Delta}^r(0,M_{vv}^2)
\right.\nonumber\\&&\left.\hspace{.5in}
+\frac{1}{6}\frac{(M_{vv}^2-M_\pi^2)}{(M_{vv}^2-M_\eta^2)}
\{3(M_{vv}^2-M_\eta^2)-(M_{vv}^2-M_\pi^2)\}
\partial_{M_{vv}^2}\bar{\Delta}^r(0,M_{vv}^2)
\right.\nonumber\\&&\left.\hspace{.5in}
-8(L_4^r-2L_6^r)(M_\pi^2+2M_K^2)-8(L_5^r-2L_8^r)M_{vv}^2
\right],\\
Z_M^{vs}&=&1+\frac{1}{2F^2}\left[
\frac{1}{3}\frac{M_\pi^2-M_\eta^2}{M_{vv}^2-M_\eta^2}
\bar{\Delta}^r(0,M_\eta^2)
+\frac{1}{3}\frac{M_{vv}^2-M_\pi^2}{M_{vv}^2-M_\eta^2}
\bar{\Delta}^r(0,M_{vv}^2)
\right.\nonumber\\&&\left.\hspace{.5in}
-8(L_4^r-2L_6^r)(M_\pi^2+2M_K^2)-8(L_5^r-2L_8^r)M_{vs}^2
\right],\\
Z_F^{vv}&=&1-\frac{1}{2F^2}\left[
2\bar{\Delta}^r(0,(M_{vv}^2+M_\pi^2)/2)+\bar{\Delta}^r(0,M_{vs}^2)
-8(L_4^r(M_\pi^2+2M_K^2)+L_5^rM_{vv}^2)
\right],\nonumber\\\\
Z_F^{vs}&=&1-\frac{1}{2F^2}\left[
\bar{\Delta}^r(0,(M_{vv}^2+M_\pi^2)/2)+\frac{1}{2}\bar{\Delta}^r(0,M_{vs}^2)
+\bar{\Delta}^r(0,M_K^2)
\right.\nonumber\\&&\left.\hspace{.5in}
+\frac{1}{3}\left(1+\frac{1}{2}\frac{M_\pi^2-M_\eta^2}{M_{vv}^2-M_\eta^2}\right)^2
\bar{\Delta}^r(0,M_\eta^2)
\right.\nonumber\\&&\left.\hspace{.5in}
+\frac{1}{3}\left\{\frac{M_{vv}^2-M_\pi^2}{M_{vv}^2-M_\eta^2}-\frac{1}{2}
-\frac{1}{4}\left(\frac{M_\pi^2-M_\eta^2}{M_{vv}^2-M_\eta^2}\right)^2\right\}
\bar{\Delta}^r(0,M_{vv}^2)
\right.\nonumber\\&&\left.\hspace{.5in}
-\frac{1}{12}\frac{(M_{vv}^2-M_\pi^2)}{(M_{vv}^2-M_\eta^2)}
\{3(M_{vv}^2-M_\eta^2)-(M_{vv}^2-M_\pi^2)\}
\partial_{M_{vv}^2}\bar{\Delta}^r(0,M_{vv}^2)
\right.\nonumber\\&&\left.\hspace{.5in}
-8(L_4^r(M_\pi^2+2M_K^2)+L_5^rM_{vs}^2)
\right],\\
\frac{\Sigma_{\rm eff}}{\Sigma} &=& 
1-\frac{1}{F^2}\left[
2\bar{\Delta}^r(0,M_\pi^2/2)+\bar{\Delta}^r(0,M_{ss}^2/2)
\right.\nonumber\\&&\left.\hspace{.5in}
-\frac{1}{3}\left\{
-\frac{(M_\pi^2-M_\eta^2)^2}{2M_\eta^4}\bar{\Delta}^r(0,M_\eta^2)
+\left(1+\frac{(M_\pi^2-M_\eta^2)^2}{2M_\eta^4}\right)
\left(-\frac{\beta_1}{\sqrt{V}}\right)
\right.\right.\nonumber\\&&\left.\left.\hspace{.5in}
+\frac{M_\pi^2(M_\pi^2-3M_\eta^2)}{2M_\eta^2}
\left(-\frac{1}{16\pi^2}\ln V^{1/2}\mu_{sub}^2-\beta_2\right)
\right\}
-16L_6^r(M_\pi^2+2M_K^2)
\right],\\
\label{eq:Sigma-2+1flavor}
\mathcal{S}_v &=& 
-\frac{1}{(\mu^{2}-\mu_v^2)^2(\mu_s^2-\mu_v^2)}
\nonumber\\
&&
\times \frac{\det \left(
\begin{array}{ccccc}
\partial_{\mu_v}K_Q(\mu_v) & I_Q(\mu_v) & I_Q(\mu) & 
\mu^{-1}I_{Q-1}(\mu) & I_Q(\mu_{s})\\
-\partial_{\mu_v} (\mu_{v} K_{Q+1}(\mu_{v})) & \mu_v I_{Q+1}(\mu_v) 
& \mu I_{Q+1}(\mu) 
& I_{Q}(\mu) &  \mu_{s}I_{Q+1}(\mu_{s}) \\
\partial_{\mu_v} (\mu_{v}^2 K_{Q+2}(\mu_{v})) & \mu_v^2I_{Q+2}(\mu_v) & 
\mu^2I_{Q+2}(\mu) 
& \mu I_{Q+1}(\mu) & \mu_{s}^2I_{Q+2}(\mu_{s})\\
-\partial_{\mu_v} (\mu_{v}^3 K_{Q+3}(\mu_{v})) & \mu_v^3I_{Q+3}(\mu_v) & 
\mu^3I_{Q+3}(\mu) 
& \mu^2I_{Q+2}(\mu) & \mu_{s}^3I_{Q+3}(\mu_{s})\\
\partial_{\mu_v} (\mu_{v}^4 K_{Q+4}(\mu_{v})) & \mu_v^4I_{Q+4}(\mu_v) & 
\mu^4I_{Q+4}(\mu) 
& \mu^3I_{Q+3}(\mu) & \mu_{s}^4I_{Q+4}(\mu_{s})
\end{array}
\right)}{\det \left(
\begin{array}{ccc}
I_Q(\mu) & \mu^{-1}I_{Q-1}(\mu) & I_Q(\mu_{s})\\
\mu I_{Q+1}(\mu) & I_{Q}(\mu) &  \mu_{s}I_{Q+1}(\mu_{s}) \\
\mu^2I_{Q+2}(\mu) & \mu I_{Q+1}(\mu) & \mu_{s}^2I_{Q+2}(\mu_{s})\\
\end{array}
\right)},\\
\label{eq:Dvv 2+1-flavor}
\mathcal{D}_{vv} &=& 
- \frac{1}{(\mu^{2}-\mu_v^2)^2(\mu_s^2-\mu_v^2)}
\nonumber\\
&&
\times \frac{\det \left(
\begin{array}{ccccc}
\partial_{\mu_v}K_Q(\mu_v) & \partial_{\mu_v} I_Q(\mu_v) & I_Q(\mu) & 
\mu^{-1}I_{Q-1}(\mu) & I_Q(\mu_{s})\\
-\partial_{\mu_v} (\mu_{v} K_{Q+1}(\mu_{v})) & \partial_{\mu_v}(\mu_v I_{Q+1}(\mu_v)) 
& \mu I_{Q+1}(\mu) 
& I_{Q}(\mu) &  \mu_{s}I_{Q+1}(\mu_{s}) \\
\partial_{\mu_v} (\mu_{v}^2 K_{Q+2}(\mu_{v})) & \partial_{\mu_v}(\mu_v^2I_{Q+2}(\mu_v)) & 
\mu^2I_{Q+2}(\mu) 
& \mu I_{Q+1}(\mu) & \mu_{s}^2I_{Q+2}(\mu_{s})\\
-\partial_{\mu_v} (\mu_{v}^3 K_{Q+3}(\mu_{v})) & \partial_{\mu_v}(\mu_v^3I_{Q+3}(\mu_v)) & 
\mu^3I_{Q+3}(\mu) 
& \mu^2I_{Q+2}(\mu) & \mu_{s}^3I_{Q+3}(\mu_{s})\\
\partial_{\mu_v} (\mu_{v}^4 K_{Q+4}(\mu_{v})) & \partial_{\mu_v}(\mu_v^4I_{Q+4}(\mu_v)) & 
\mu^4I_{Q+4}(\mu) 
& \mu^3I_{Q+3}(\mu) & \mu_{s}^4I_{Q+4}(\mu_{s})
\end{array}
\right)}{\det \left(
\begin{array}{ccc}
I_Q(\mu) & \mu^{-1}I_{Q-1}(\mu) & I_Q(\mu_{s})\\
\mu I_{Q+1}(\mu) & I_{Q}(\mu) &  \mu_{s}I_{Q+1}(\mu_{s}) \\
\mu^2I_{Q+2}(\mu) & \mu I_{Q+1}(\mu) & \mu_{s}^2I_{Q+2}(\mu_{s})\\
\end{array}
\right)}
\nonumber\\&&
+ \left(\frac{4\mu_v}{\mu^2-\mu_v^2}+\frac{2\mu_v}{\mu_s^2-\mu_v^2}\right) 
\mathcal{S}_v.
\end{eqnarray}
Here we have used explicit expressions for $\bar{G}(0,M_1^2,M_2^2)$'s
shown in Ref.~\cite{Aoki:2009mx} and
\begin{eqnarray}
\lim_{M\to 0}\bar{\Delta}^r(0,M^2) &=& -\frac{\beta_1}{\sqrt{V}},\\
\lim_{M\to 0}\partial_{M^2}\bar{\Delta}^r(0,M^2) 
&=& -\frac{1}{16\pi^2}\ln V^{1/2}\mu_{sub}^2 -\beta_2.
\end{eqnarray}

For $\mathcal{S}_v$ and $\mathcal{D}_{vv}$ 
at degenerate up and down quark masses,
we have used an expansion
$(\mu+\Delta \mu)^\alpha I_{\alpha}(\mu+\Delta \mu)=
\mu^\alpha I_{\alpha}(\mu)+\mu\Delta 
\mu [\mu^{\alpha-1} I_{\alpha-1}(\mu)]+{\cal O}((\Delta \mu)^2)$ 
for any $\alpha$, and a similar expansion for $K_\nu$'s.
Note that $\mathcal{S}^{\rm eff}$ and $\mathcal{D}^{\rm eff}$
are obtained by simply replacing $\Sigma$ with $\Sigma_{\rm eff}$
in the above formulas.

\subsection{When $M_{vv}\ll M_\pi$}

In Eq.~(\ref{eq:r2and2+1}), we have neglected
a term proportional to $M_{vv}^2-M_\pi^2$.
One might, however, encounter the case where
one wants to reduce the valence quark mass
to the very vicinity of the chiral limit while
keeping the physical pion mass at the $p$ regime.
In such a case, a partial quenching artifact
is enhanced as a double-pole contribution
and one has to add the following contributions 
to the pion correlator, 
\begin{eqnarray}
\Delta \mathcal{\pi}(t,m_{v})_Q &=&
-F_{PP}^Q \partial_{M_{vv}^2}\left[
\frac{\cosh (M_{vv}(t-T/2))}{2M_{vv}\sinh(M_{vv}T/2)}-\frac{1}{M_{vv}^2T}
\right],
\end{eqnarray}
where
\begin{eqnarray}
F_{PP}^Q &=& 
\left\{
\begin{array}{ll}
{\displaystyle \frac{\Sigma^2}{F^2}
\left(\frac{\partial \mathcal{S}_v}{\partial \mu_v}\right)
(M_{vv}^2-M_\pi^2)}&(N_f=2),\\\\
{\displaystyle\frac{\Sigma^2}{F^2}
\left(\frac{\partial \mathcal{S}_v}{\partial \mu_v}\right)
(M_{vv}^2-M_\pi^2)\left(1-\frac{1}{3}
\frac{M_{vv}^2-M_\pi^2}{M_{vv}^2-M_\eta^2}\right)}
&(N_f=2+1).
\end{array}
\right. 
\end{eqnarray}

\subsection{Masses and decay constants}

\begin{figure*}[btp]
  \centering
  \includegraphics{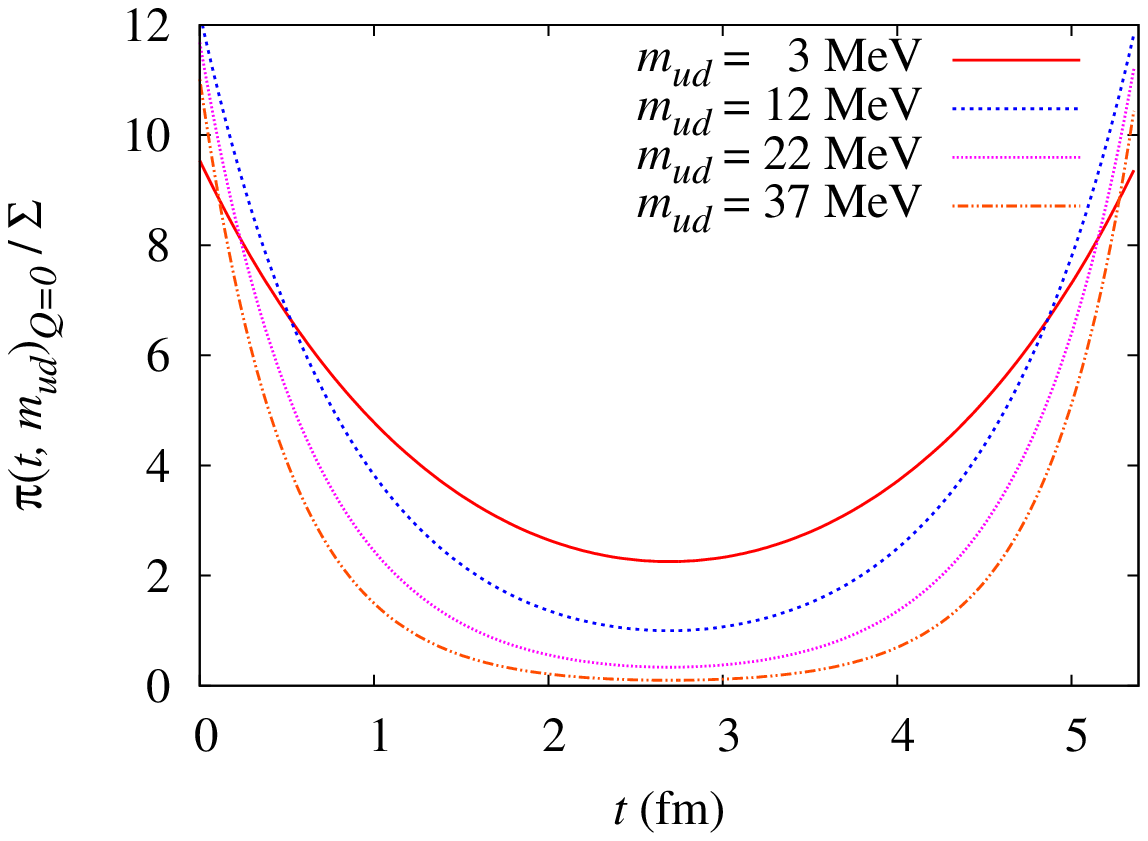}
  \caption{
The $N_f=2+1$ ChPT prediction for the 
pion correlator $\mathcal{\pi}(t,m_{v})_Q$ 
(normalized by $\Sigma$) 
at $Q=0$ and $m_v=m_{ud}$.  
The finite periodic box size is
$V=L^3T=(1.8~\mbox{fm})^3\times (5.4~\mbox{fm})$.
We use $m_s=111$ MeV, $\Sigma^{\overline{\rm MS}}(2\;\mbox{GeV})
=[234\;\mbox{MeV}]^3$, $F=71\;\mbox{MeV}$, 
$L_4^r(770\;\mbox{MeV})=0.0$, $L_5^r(770\;\mbox{MeV})=2.2\times 10^{-3}$,
 $L_6^r(770\;\mbox{MeV})=0.0$ and $L_8^r(770\;\mbox{MeV})=1.1\times 10^{-3}$
as the inputs.
}
  \label{fig:pioncorr}
  \centering
  \includegraphics{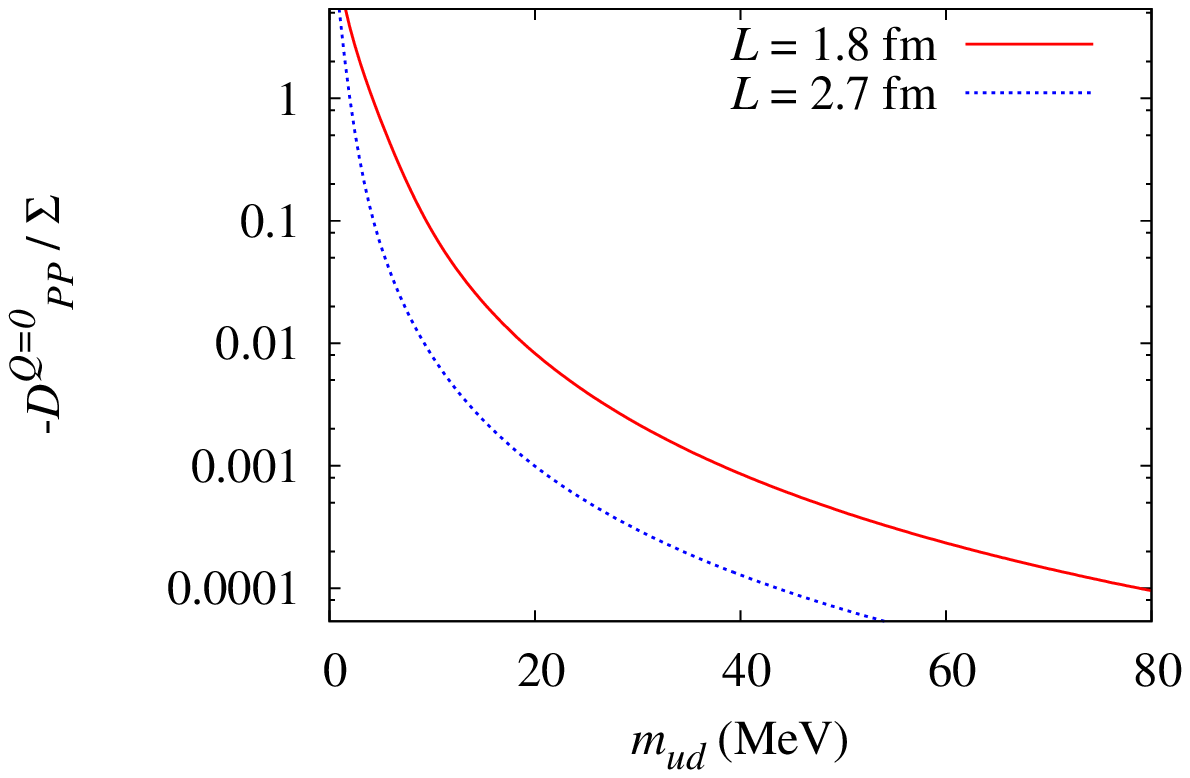}
  \caption{
The quark mass 
$m_{ud}(=m_{v})$ dependence 
of $-D^Q_{PP}$ (normalized by $\Sigma$) 
at $Q=0$. 
The same inputs as Fig.~\ref{fig:pioncorr}  are used.
}
  \label{fig:DPP}
\end{figure*}

In this subsection we demonstrate how to extract
the masses and decay constants of the pions (and kaons) from lattice QCD data 
using our formula. 
We plot in Fig.~\ref{fig:pioncorr} the pion correlator
Eq.~(\ref{eq:pion}) (normalized by $\Sigma$)
at several different quark masses.
We take $m_{ud}=m_v$ in all cases.
In the plot, the strange quark mass is fixed at $m_s=111$ MeV,
and the topological charge is fixed at $Q=0$.
We choose the finite box size as
$V=L^3T=(1.8\;\mbox{fm})^3\times (5.4\;\mbox{fm})$
and the boundary condition is periodic in all directions.
For the inputs, we use one of the latest lattice QCD results for 
the chiral condensate and the pion decay constant,  
$\Sigma
=[234\;\mbox{MeV}]^3$ (in the $\overline{\rm MS}$ scheme at 2 GeV)
and $F=71\;\mbox{MeV}$  from Ref.~\cite{Fukaya:2010na}.
For the other low energy constants, phenomenological estimates 
from Ref.~\cite{Gasser:1983yg},
$L_4^r(770\;\mbox{MeV})=0.0$, $L_5^r(770\;\mbox{MeV})=2.2\times 10^{-3}$,
 $L_6^r(770\;\mbox{MeV})=0.0$, and $L_8^r(770\;\mbox{MeV})=1.1\times 10^{-3}$
are used.

As the first step of the analysis, 
one should identify the presence (or absence) of
the constant term $D_{PP}^Q$, which is a signal of entering
(or leaving) the $\epsilon$ regime.
As shown in Fig.~\ref{fig:DPP}, it is a rapidly 
decreasing function of the quark mass.
Since this constant comes from the zero-mode part,
it is essentially controlled by
the chiral condensate.
Using lattice QCD data for $\Sigma$ (or $\Sigma_{\rm eff}$) 
or taking time derivative of the correlator,
$D_{PP}^Q$ can be subtracted.

\begin{figure*}[tbhp]
  \centering
  \includegraphics{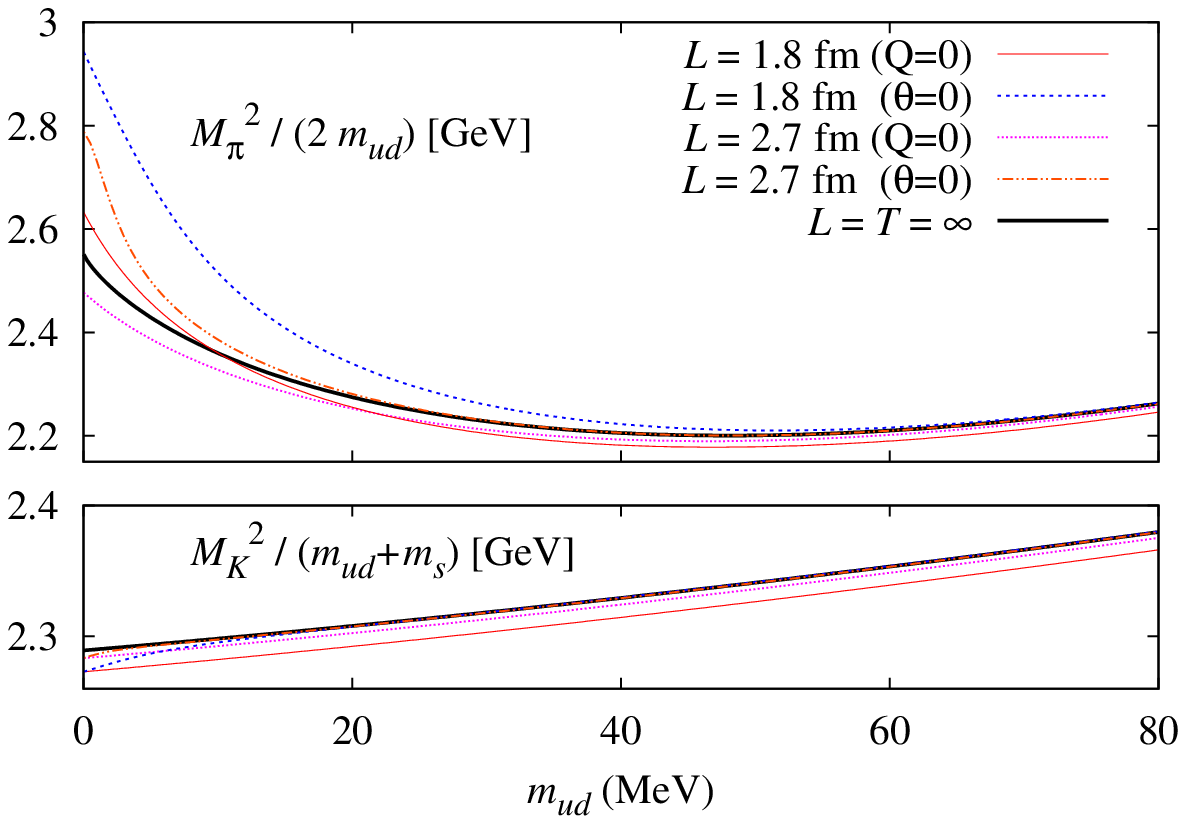}
  \caption{
The up-down quark mass $m_{ud}(=m_{v})$ dependence of 
$(M_{vv}^{Q=0/\theta=0})^2/(2m_{ud})$ (upper panel) and 
$(M_{vs}^{Q=0/\theta=0})^2/(m_{ud}+m_s)$ (lower) is plotted
at different volume sizes.
The same inputs as Fig.~\ref{fig:pioncorr}  are used.
}
  \label{fig:M}
  \centering
  \includegraphics{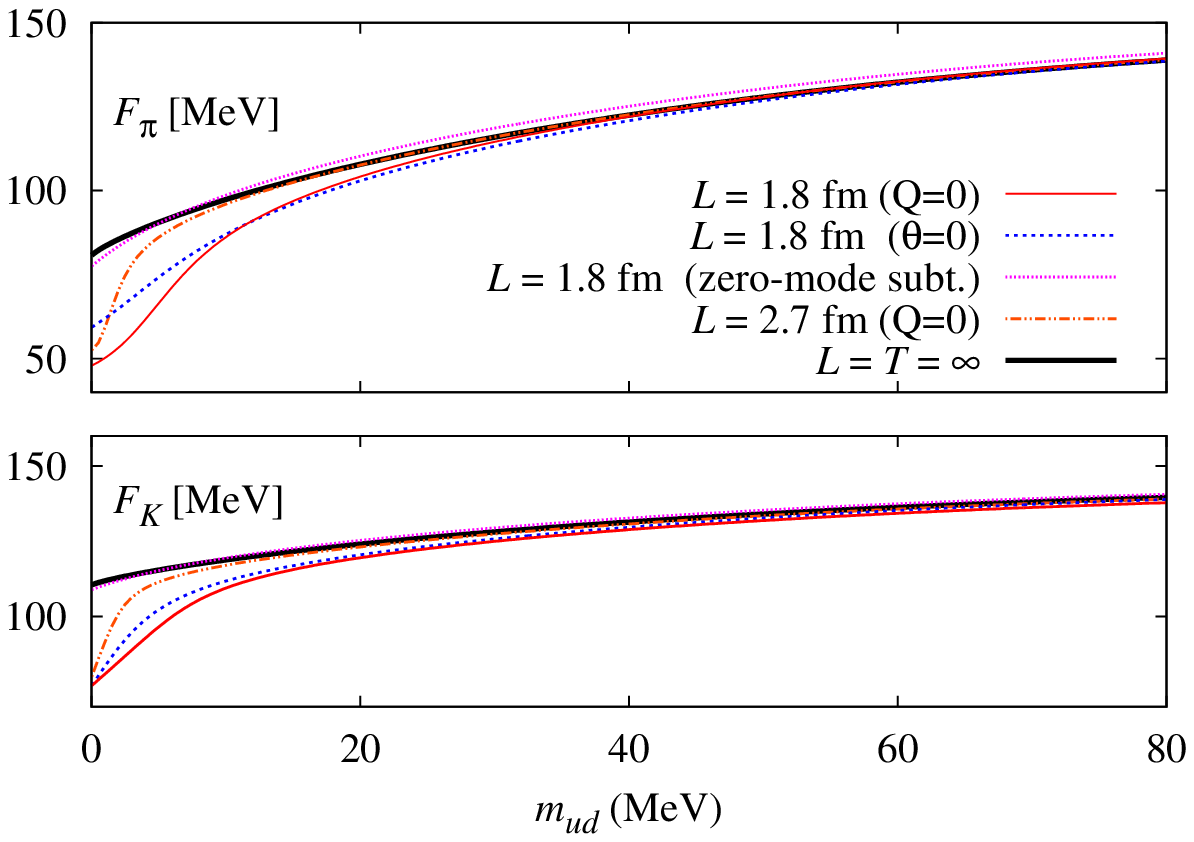}
  \caption{
$m_{ud}(=m_v)$ dependence of 
the pion (upper panel) and kaon (lower) 
decay constants $F_\pi^{Q=0/\theta=0}$ and $F_K^{Q=0/ \theta=0}$
at different volume sizes.
The curves with the index ``zero-mode subt.'' denote
$F_\pi^\prime$ or $F_K^\prime$.
See the text for the notation.
The same inputs as Fig.~\ref{fig:pioncorr}  are used.
}
  \label{fig:F}
\end{figure*}

Next, from the remaining $\cosh$ function part,
the meson masses can be determined.
In Fig.~\ref{fig:M}, we plot the quark mass 
dependence of the pion mass squared divided by the quark mass: 
$(M_{vv}^{Q=0/\theta=0})^2/(2 m_{ud})$ and 
that for the kaon mass: $(M_{vs}^{Q=0/\theta=0})^2/(m_{ud}+m_s)$.
Here the same inputs shown above are used.
The $\theta=0$ results here and in the following are calculated 
via Eq.~(\ref{eq:sumQ}) truncating the sum at $|Q|=20$,
which already shows a good convergence.
For the pion mass, $10$--20\% deviation from the 
infinite $V$ result (thick curves) is found near the 
chiral limit while the kaon mass 
suggests only $\sim$1\%  finite volume effects.
Note that there is no contribution from the zero-mode
to the meson masses at $Q=0$ (See Eq.(\ref{eq:MQ}).).

Finally let us discuss how to determine
the pion decay constant from the
coefficient $C_{PP}^Q$.
It is not difficult to check that 
a naive conventional definition 
$F^Q_{\pi} \equiv \sqrt{ 4 m_v^2 C_{PP}^Q/(M_{vv}^Q)^4}$
or its counterpart in the $\theta=0$ vacuum
$F^{\theta =0}_{\pi}\equiv (F_{\pi}^Q)_{\theta=0}$
actually leads to the right infinite volume limit 
$F_\pi=FZ_F^{vv}|_{V=\infty}$ as $V$ increases.
It is also the case for the kaon decay constant: 
$F^Q_K \equiv \sqrt{  (m_v+m_s)^2 E_{PP}^Q/(M_{vs}^Q)^4}$ (or $F^{\theta =0}_K$)
converges to the infinite volume limit of $F_K=FZ_F^{vs}|_{V=\infty}$.
Note however that the curves in Fig.~\ref{fig:F} show
a considerable deviation ($\sim$ 50\%) 
as the quark mass is reduced, which is 
a typical consequence of the non-perturbative zero-mode integrals.
Unlike the meson masses, 
not only the pion decay constant but also the 
kaon decay constant receives
a large contribution from the zero-mode.
These zero-mode integrals are again controlled by
the chiral condensate, and therefore
one should in principle be able to 
subtract this part using lattice QCD data for $\Sigma$ (or $\Sigma_{\rm eff}$).
Once the zero-mode part, 
$\mathcal{D}_{vv}^{\rm eff}-1
+\frac{Q^2}{(\mu_{v}^{\rm eff})^2} -\frac{\partial \mathcal{S}_v^{\rm eff}}{\partial \mu^{\rm eff}_v}$
or $\mathcal{S}_{v}^{\rm eff}
\left(1-\frac{2}{\mu_s+\mu}-\frac{1}{2\mu_s}+\frac{Q^2}{2\mu_s^2}\right)
+\frac{Q^2}{\mu_{v}\mu_{s}}-1 $,
 is subtracted, one obtains
$F_\pi^\prime \equiv F Z_F^{vv}$ or $F_K^\prime \equiv F Z_F^{vs}$,
which have a much milder volume dependence (at most a few \% level) 
as shown by the dotted curves in Fig.~\ref{fig:F}.

\begin{figure*}[tbp]
  \centering
  \includegraphics{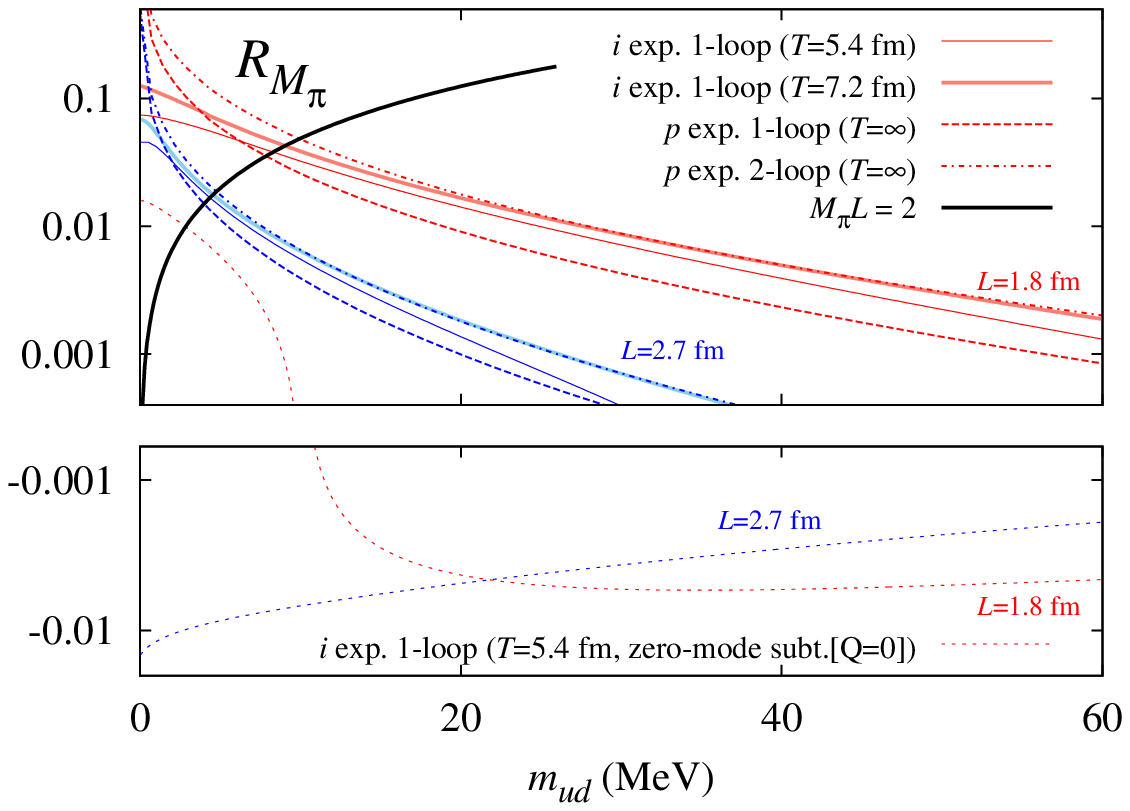}
  \includegraphics{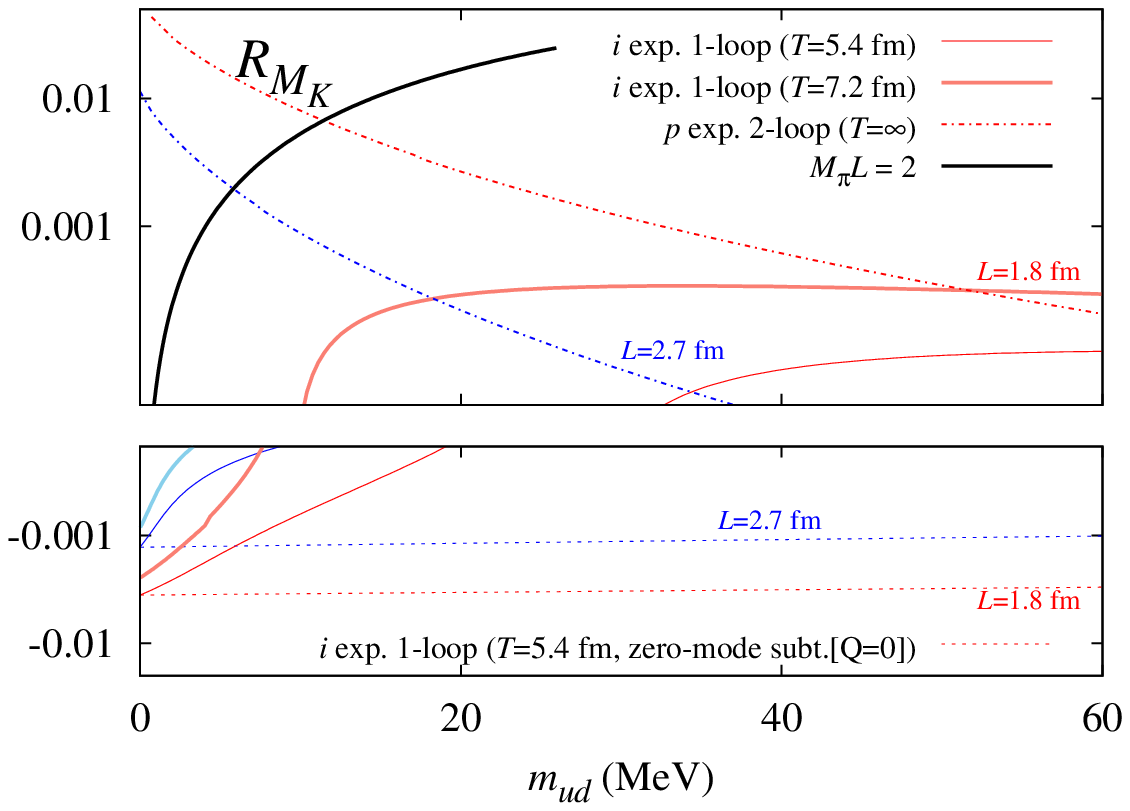}
  \caption{
Comparison with the $p$ expansion results 
\'{a} la L\"uscher formula \cite{Luscher:1985dn}.
Our new ChPT calculation ($i$ exp.)
and the $p$ expansion ($p$ exp.) result from Ref.~\cite{Colangelo:2005gd} for 
$R_{M_\pi}$ (top) and $R_{M_K}$ (bottom) are drawn 
(note that one-loop correction in the $p$ expansion on $R_{M_K}$ is zero).
The same inputs as Fig.~\ref{fig:pioncorr} 
and those given in Ref.~\cite{Colangelo:2005gd} for
the other higher LECs are used.
The $M_\pi L=2$ (thick) curve is drawn 
using the two-loop result. 
}
  \label{fig:RM}
\end{figure*}

\begin{figure*}[tbp]
  \centering
  \includegraphics{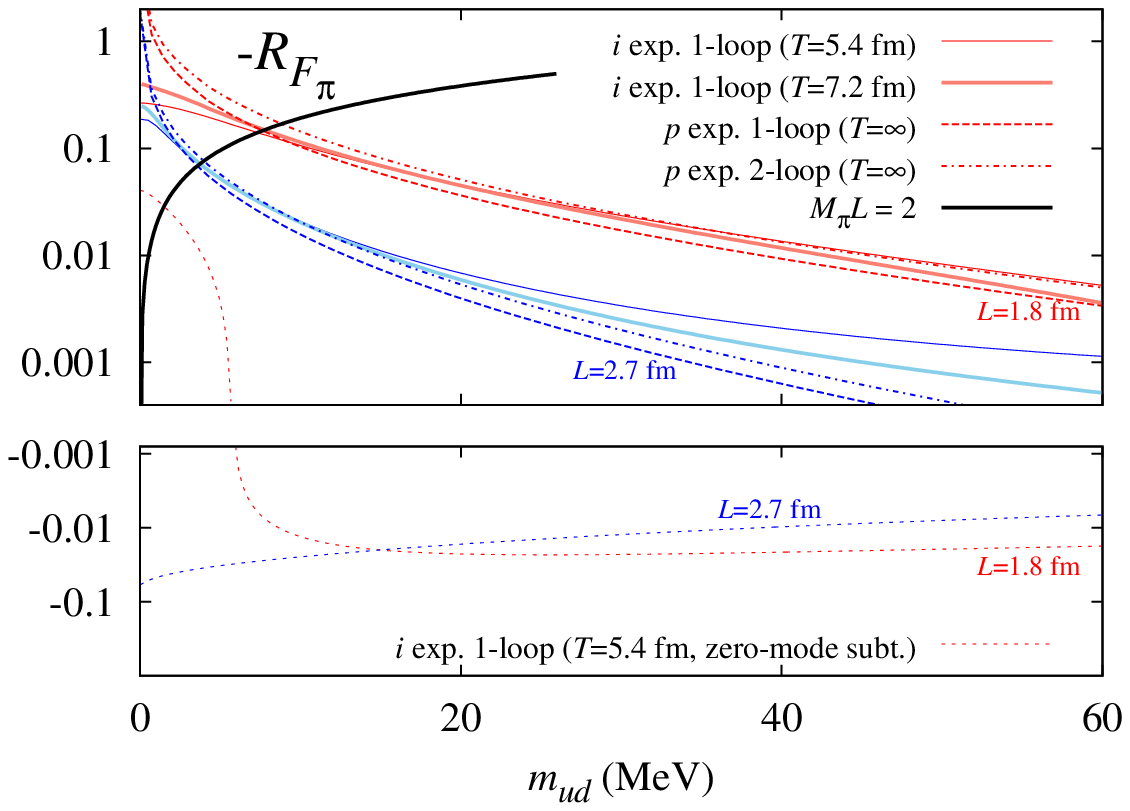}
  \includegraphics{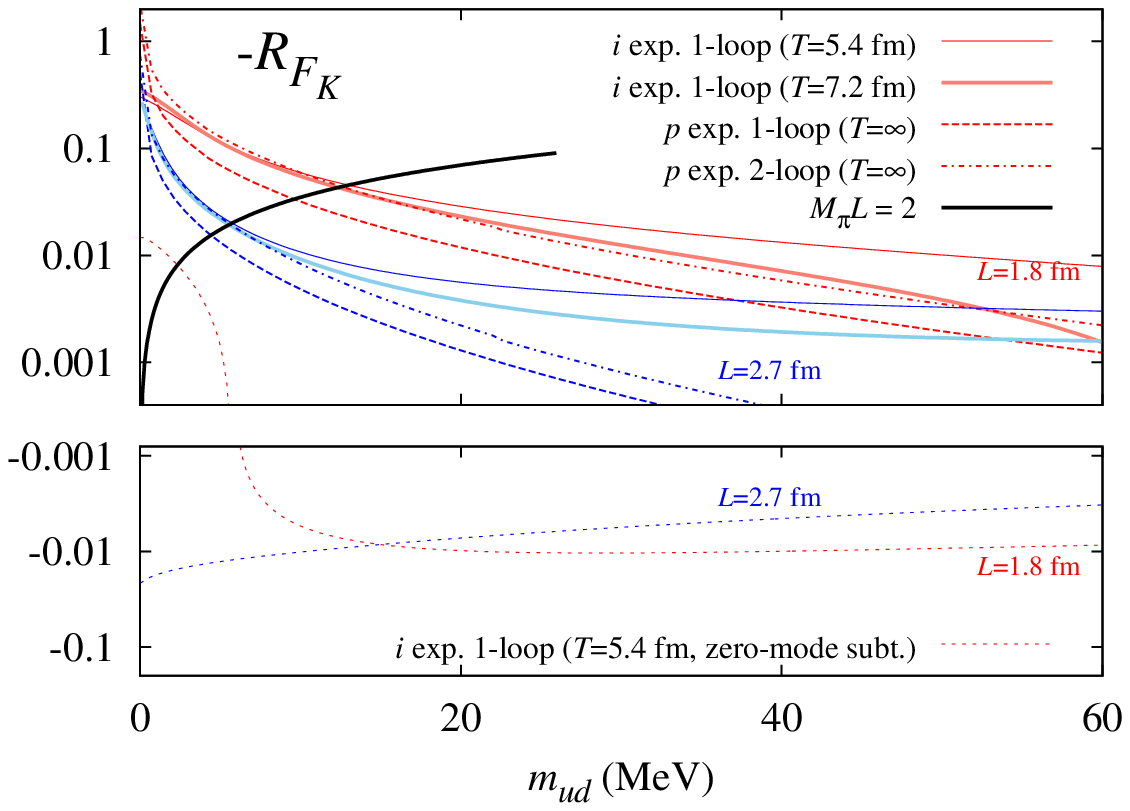}
  \caption{
Comparison with the $p$ expansion results \cite{Colangelo:2005gd} for 
$-R_{F_\pi}$ (top) and $-R_{F_K}$ (bottom).
}
  \label{fig:RF}
\end{figure*}

We emphasize that the accuracy of our calculation is
NLO even though the zero-mode contribution
is partly treated to all-order.
It is interesting to compare our results with the conventional
finite volume formulas in the $p$ expansion
since higher order loop calculations 
are available \'{a} la L\"uscher formula \cite{Luscher:1985dn} for the latter.
In Figs.~\ref{fig:RM} and \ref{fig:RF}, we plot
our results for
\begin{eqnarray}
R_{M_{\pi/K}} \equiv \frac{M^{Q/\theta=0}_{\pi/K}(L)}{M_{\pi/K}(L=\infty)}-1,
\;\;\;\;\;
R_{F_{\pi/K}} \equiv \frac{F^{Q/\theta=0}_{\pi/K}(L)}{F_{\pi/K}(L=\infty)}-1,
\end{eqnarray}
comparing with those in the two-loop (and one-loop) calculations 
in the $p$ expansion
by Colangelo {\it et al}. \cite{Colangelo:2005gd}.
The same inputs for $\Sigma$, $F$ and $L_i$'s above are used. 
For the other higher order LECs, the values given in 
\cite{Colangelo:2005gd} are used.

Our formula at one-loop 
(denoted by $i$ exp.) in the $\theta=0$ vacuum is 
drawn by the solid ($T=$5.4 fm) and
thick ($T=$7.2 fm) curves while
the dotted curves ($T=$5.4 fm) show the results from which 
the zero-mode contribution is subtracted.
Note that even in the region $M_\pi L<2$, 
our formulas are finite while the $p$ expansion
(dashed curves)
results show an unphysical divergence.
For $M_\pi L>2$, on the other hand,
we observe that our result is consistent with the $p$ expansion.
It is, in particular, remarkable that our one-loop result
is closer to the two-loop formula rather than one-loop in the $p$ expansion.
In order to understand whether this is a just coincidence
or can be explained by the effect of the zero-mode resummation,
a further study 
in the limit of $T\to \infty$, which enters another
regime (the $\delta$ regime 
\cite{Leutwyler:1987ak, Hasenfratz:1993vf, 
Hasenfratz:2009mp, Bietenholz:2010az, Weingart:2010yv, 
Niedermayer:2010mx, Bietenholz:2011ia}), is needed.\\

We have observed that, as the quark masses decrease, 
the pseudoscalar correlator in a finite volume is largely distorted from 
the form in the infinite volume limit because of the zero-momentum mode fluctuation.
By a careful removal of its contribution using the ChPT formulas, however, 
we can obtain a milder volume dependence, which makes 
it  possible to extract the $V\to \infty$ limit
of the meson masses or decay constants.

\clearpage
\section{A short-cut prescription}
\label{sec:discussion}

We have performed a complete calculation to obtain
the general form of 
the pseudoscalar correlation function
in Eq.~(\ref{eq:PPd3x}), which contains a
conventional $\cosh$ function as well as
a constant term and a contribution 
from 3-particle states.

It is no surprise that the
constant term appears
since the correlator in the conventional $p$ regime
shows an unphysical 
infra-red divergence in the chiral limit.
To remove this divergence, 
the zero-mode or the constant mode contribution
is indispensable.

With this observation we find that the result in
Eq.~(\ref{eq:PPd3x}) is obtained by an easier prescription below.
Starting from the conventional $p$ expansion formula in 
Eq.~(\ref{eq:PPptheta}),
\begin{enumerate}
\item Replace the $Z$ factors with those from which 
the zero-mode contribution is subtracted, namely, 
$[Z^{ij}_F]_p$ and $[Z^{ij}_M]_p$ with $Z^{ij}_F$ and $Z^{ij}_M$.
\item Replace 
\begin{equation}
\frac{\cosh (M_{ij}^{\theta=0}(t-T/2))}
{M_{ij}^{\theta=0}\sinh(M_{ij}^{\theta=0}T/2)}\;\;\;\mbox{with}\;\;\; 
\frac{\cosh (M_{ij}^Q(t-T/2))}
{M_{ij}^Q\sinh(M_{ij}^Q T/2)}-\frac{2}{(M_{ij}^Q)^2T}.
\end{equation}
\item Multiply  a factor coming from the exact 
zero-mode integrals, which can be read off
from the coefficient of the $t$ dependent term or 
the $2Th_1(t/T)$ term in the $\epsilon$ expansion result. 
In the case of Eq.~(\ref{eq:PPd3x}), it is 
$\frac{1}{2}
\left(1+\mathcal{D}^{\rm eff}_{12}
+\frac{Q^2}{\mu^{\rm eff}_1\mu^{\rm eff}_2}\right)$ obtained 
from Eq.~(\ref{eq:PPepsilon}).
Note that the NLO condensate $\Sigma_{\rm eff}$, which contains
chiral-log terms, should be used instead of the bare value $\Sigma$. 
\item Add the constant and $r_{12}(t)$ terms if 
they exist in the $\epsilon$ expansion.  
\end{enumerate}

In fact, in a similar prescription, it is not difficult to
obtain (a conjecture for) 
the axialvector-pseudoscalar and axialvector-axialvector correlators:
\begin{eqnarray}
\label{eq:AP}
\mathcal{AP}(t,m_{1},m_{2})_Q &\equiv& 
\int d^3 x \langle A_0(x)P(0)\rangle_Q \hspace{-2in} 
\nonumber\\&&\hspace{-1in}
=\frac{\Sigma(Z_F^{12} Z_M^{12})^2}{2}
\left[\left(1+\mathcal{D}^{\rm eff}_{12}
+\frac{Q^2}{\mu^{\rm eff}_1\mu^{\rm eff}_2}\right)
\left/\left(1+\frac{Q^2}{2\mu_1\mu_2}\right)\right.\right]
\frac{\sinh (M_{12}^Q (t-T/2))}{\sinh (M_{12}^Q T/2)}
\nonumber\\&&\hspace{-0.8in}
+\Sigma(Z_F^{12} Z_M^{12})^2\left[
\left(\mathcal{S}^{\rm eff}_1+\mathcal{S}^{\rm eff}_2\right)
-\left(1+\mathcal{D}^{\rm eff}_{12}
+\frac{Q^2}{\mu^{\rm eff}_1\mu^{\rm eff}_2}\right)
\left/\left(1+\frac{Q^2}{2\mu_1\mu_2}\right)\right.\right]
\left(\frac{t}{T}-\frac{1}{2}\right)
\nonumber\\&&\hspace{-0.8in}
-M_{12}^2\Sigma\frac{\mathcal{S}_1-\mathcal{S}_2}{\mu_1-\mu_2}
\frac{\partial}{\partial t}
(\partial_{M_{11}^2}r_{12} (t)+\partial_{M_{22}^2}r_{12} (t)),\\
\label{eq:AA}
\mathcal{AA}(t,m_{1},m_{2})_Q &\equiv& 
\int d^3 x \langle A_0(x)A_0(0)\rangle_Q \hspace{-2in} 
\nonumber\\
&=&
-\frac{\Sigma (Z_F^{12}Z_M^{12})^2}{2}
(m_1+m_2)\left(\mathcal{S}^{\rm eff}_1+\mathcal{S}^{\rm eff}_2\right)
\frac{\cosh (M_{12}^Q(t-T/2))}{M_{12}^Q\sinh(M_{12}^QT/2)}
\nonumber\\&&
+\frac{(F Z^{12}_F)^2}{T}\left[
\left(\mathcal{S}^{\rm eff}_1+\mathcal{S}^{\rm eff}_2\right)
\left/\left(1+\frac{Q^2}{2\mu_1\mu_2}\right)\right.
-\left(1+\mathcal{D}_{12}^{\rm eff}
-\frac{Q^2}{\mu_1^{\rm eff}\mu_2^{\rm eff}}\right)
\right]
\nonumber\\&&
+\frac{T}{2V}\left(1-\mathcal{D}_{12}^{\rm eff}
+\frac{Q^2}{\mu_1^{\rm eff}\mu_2^{\rm eff}}\right)
\sum_i^{N_f} (\bar{k}_{00}(M_{1i}^2)+\bar{k}_{00}(M_{2i}^2)),
\end{eqnarray}
where 
\begin{eqnarray}
\bar{k}_{00}(M^2) &\equiv& 
\sum_{{\bf q}=(p_1,p_2,p_3)} 
\frac{-1}{4\sinh^2(\sqrt{|{\bf q}|^2+M^2}\;T/2)}+\frac{1}{M^2T^2},
\end{eqnarray}
which is UV finite (and of course IR finite as well)
and can be thus numerically evaluated.

We confirm that Eqs.~(\ref{eq:AP}) and Eq.~(\ref{eq:AA})
indeed converge to those in the $p$ expansion \cite{Aoki:2009mx}
for the larger masses
and those in the $\epsilon$ expansion 
\cite{Damgaard:2007ep, Bernardoni:2008ei}
near the chiral limit.
The above prescription thus achieves at least  
a smooth interpolation between the $\epsilon$ and $p$ regimes.
Note  that the $\epsilon$ regime result is not found in the literature for the $\mathcal{AP}(t,m_{1},m_{2})_Q$ correlator.
We present in 
Appendix~\ref{app:APepsilon} our own calculation.

Furthermore, we find a more non-trivial evidence 
that supports our prescription:
the axial Ward-Takahashi identities 
\begin{eqnarray}
\frac{\partial}{\partial t}\mathcal{AP}(t,m_{1},m_{2})_Q
&=& (m_1+m_2)\mathcal{PP}(t,m_{1},m_{2})_Q
\nonumber\\
&& \hspace{-1in}=\frac{\Sigma(Z_F^{12} Z_M^{12})^2}{2}
\left[\left(1+\mathcal{D}^{\rm eff}_{12}
+\frac{Q^2}{\mu^{\rm eff}_1\mu^{\rm eff}_2}\right)
\left/\left(1+\frac{Q^2}{2\mu_1\mu_2}\right)\right.\right]
\frac{M_{12}^Q\cosh (M_{12}^Q (t-T/2))}{\sinh (M_{12}^Q T/2)}
\nonumber\\&&\hspace{-0.8in}
+\left.\frac{\Sigma(Z_F^{12} Z_M^{12})^2}{T}\right[
\left(\mathcal{S}^{\rm eff}_1+\mathcal{S}^{\rm eff}_2\right)
-\left.\left(1+\mathcal{D}^{\rm eff}_{12}
+\frac{Q^2}{\mu^{\rm eff}_1\mu^{\rm eff}_2}\right)
\left/\left(1+\frac{Q^2}{2\mu_1\mu_2}\right)\right.\right]
\nonumber\\&&\hspace{-0.8in}
-2\Sigma\frac{\mathcal{S}_1-\mathcal{S}_2}{\mu_1-\mu_2}
M_{12}^2r_{12}(t)+{\cal O}(p^4),
\end{eqnarray}
and 
\begin{eqnarray}
-\frac{\partial}{\partial t}\mathcal{AA}(t,m_{1},m_{2})_Q
&=& (m_1+m_2)\mathcal{AP}(t,m_{1},m_{2})_Q
\nonumber\\
&=& \Sigma (Z_F^{12}Z_M^{12})^2
(m_1+m_2)\frac{\mathcal{S}^{\rm eff}_1+\mathcal{S}^{\rm eff}_2}{2}
\frac{\sinh (M_{12}^Q(t-T/2))}{\sinh(M_{12}^QT/2)}+{\cal O}(p^4),
\nonumber\\
\end{eqnarray}
are precisely satisfied. Here we have used 
\begin{eqnarray}
\left[
\left(\mathcal{S}^{\rm eff}_1+\mathcal{S}^{\rm eff}_2\right)
-\left(1+\mathcal{D}^{\rm eff}_{12}
+\frac{Q^2}{\mu^{\rm eff}_1\mu^{\rm eff}_2}\right)
\left/\left(1+\frac{Q^2}{2\mu_1\mu_2}\right)\right.\right]
\left(\frac{t}{T}-\frac{1}{2}\right)
\hspace{-4.7in}&&\nonumber\\
&=&
\left[\left(\mathcal{S}^{\rm eff}_1+\mathcal{S}^{\rm eff}_2\right)
-\left(1+\mathcal{D}^{\rm eff}_{12}
+\frac{Q^2}{\mu^{\rm eff}_1\mu^{\rm eff}_2}\right)
\left/\left(1+\frac{Q^2}{2\mu_1\mu_2}\right)\right.
\right]\frac{\sinh (M_{12}^Q (t-T/2))}{2\sinh (M_{12}^Q T/2)},
\end{eqnarray}
which is valid up to a higher order contribution near the chiral limit,
and
\begin{eqnarray}
\frac{\partial^2}{\partial t^2}(\partial_{M_{ii}^2}r_{12} (t))
&=& r_{12} (t)+{\cal O}(M_{ii}^2)\;\;\;\;\;(i=1,2).
\end{eqnarray}
Our results in Eqs.(\ref{eq:PPd3x}),
(\ref{eq:AP}) and (\ref{eq:AA})  
not only smoothly connect the $\epsilon$ and $p$ regimes
but also keep the symmetry of the theory
even in the intermediate region.
\section{Conclusion}
\label{sec:conclusion}

With the new perturbative scheme of ChPT proposed in 
Ref.~\cite{Damgaard:2008zs}, we have calculated
the two-point correlation function in the pseudoscalar channel.
The counting rule for the computation is essentially 
the same as in the conventional $p$ expansion 
(except for the additional rule for 
the mixing term of the zero and non-zero modes)
while some of the zero-mode integrals are performed non-perturbatively  as in the
$\epsilon$ expansion.

As seen in Eqs.(\ref{eq:PPd3x}) and Eq.(\ref{eq:pion}),
the correlator is expressed by a hyperbolic cosine function of time $t$
plus an additional constant term as well as a non-trivial 
contribution from three-particle states, which smoothly interpolates
the $p$ regime results and those in the $\epsilon$ regime. 

The presence of the constant term in the correlator was known as 
a remarkable feature of the $\epsilon$ expansion. 
We have found that this constant plays an essential role in 
canceling the unphysical divergence coming from the 
$\cosh$ term in the $p$ expansion
and keep the correlator always IR finite.

Giving examples for the $N_f=2$ and 2+1 theories,
we have proposed a new method of determining the meson masses
and decay constants from lattice QCD data obtained  in a finite volume.
Once one has a good control of the chiral condensate $\Sigma$,
and therefore, of the non-trivial coefficients $C_{PP}^Q$, 
$D_{PP}^Q$ and $E_{PP}^Q$ in the correlators Eqs.~(\ref{eq:pion})
and (\ref{eq:kaon}), the zero-mode contributions can be subtracted
and the remaining meson masses (see Fig.~\ref{fig:M}) 
and decay constants (Fig.~\ref{fig:F}) show a much milder
volume dependence.
Our results will be useful to precisely estimate 
the finite volume effects in lattice QCD data for 
the pion mass $M_\pi$ and kaon mass $M_K$, 
as well as their decay constants $F_\pi$ and $F_K$. 

From our calculation we have found a short-cut prescription 
as shown in Section \ref{sec:discussion}.
According to this greatly simplified scheme, 
we have derived  the axialvector-pseudoscalar 
and axialvector-axialvector correlators.
It turned out that these results not only give a
smooth interpolation between the $\epsilon$ and $p$ 
regimes but also keep the axial Ward-Takahashi 
identities at an arbitrary choice of quark masses.
It will be important to check if this simplified 
prescription is valid for the other quantities
like three or four point functions.


\begin{acknowledgments}
  The authors thank the members of JLQCD and TWQCD Collaborations
  for their encouragement to this study.
  HF thanks P.~H.~Damgaard for helpful discussions.
  The work of SA is supported in part by the Grant-in-Aid of the
  Japanese Ministry of Education, Sciences and Technology, Sports and Culture (No. 20340047) and by Grant-in-Aid for Scientific Research on Innovative Areas (No.2004: 20105001, 20105003).
\end{acknowledgments}

\appendix

\section{$\xi$ correlators in finite volume}
\label{app:xi}

Integrals over $\xi$ fields are expressed by
$\bar{\Delta}(x,M^2)$ and $\bar{G}(x,M_1^2,M_2^2)$
defined by Eqs.~(\ref{eq:Deltabar}) and (\ref{eq:Gbar}).
Here we summarize useful formulas in the
calculation of meson correlators.

We first note that even simple (three-dimensional) integrals
and derivatives of them
have unusual forms like 
\begin{eqnarray}
\int d^3 x \;\bar{\Delta}(x,M^2) &=& 
\frac{\cosh(M(t-T/2))}{2M \sinh(MT/2)}-\frac{1}{M^2T}, \label{eq:space_int}\\
\partial_\mu^2\bar{\Delta}(x,M^2)&=& M^2\bar{\Delta}(x,M^2)+\frac{1}{V},
\end{eqnarray} 
due to absence of the zero-mode.

For the ${\cal O}(S_I^{(1)})$ contribution, we need
\begin{eqnarray}
\label{eq:DD}
\int d^4 y \bar{\Delta}(x-y,M_1^2)\bar{\Delta}(y,M_2^2)
&=& \frac{1}{V}\sum_{p\neq 0}\frac{e^{ipx}}{(p^2+M_1^2)(p^2+M_2^2)}
\nonumber\\
&=& \frac{1}{M_2^2-M_1^2}
\left(\bar{\Delta}(x,M_1^2)-\bar{\Delta}(x,M_2^2)\right).
\end{eqnarray}
which becomes 
$-\partial_{M^2}\bar{\Delta}(x,M^2)|_{M^2=M_1^2}$ in the limit $M^2_1=M^2_2$.

In the same way,
\begin{eqnarray}
\label{eq:DG}
\int d^4 y \bar{\Delta}(x-y,M_1^2)\bar{G}(y,M_2^2,M_3^2)
&=& \frac{1}{M_2^2-M_1^2}
\left(\bar{G}(x,M_1^2,M_3^2)-\bar{G}(x,M_2^2,M_3^2)\right),
\end{eqnarray}
which can be expressed in two different ways:
\begin{eqnarray}
\label{eq:DG2}
&=& \frac{1}{M_3^2-M_2^2}
\left(\bar{G}(x,M_1^2,M_2^2)-\bar{G}(x,M_1^2,M_3^2)\right),
\nonumber\\
&=& \frac{1}{M_3^2-M_1^2}
\left(\bar{G}(x,M_1^2,M_2^2)-\bar{G}(x,M_2^2,M_3^2)\right).
\end{eqnarray}

For ${\cal O}((S_I^{(1)})^2)$ contributions, we use
\begin{eqnarray}
\int d^4 y\, d^4z\, \bar{\Delta}(x-y,M_1^2)\bar{\Delta}(y-z,M_2^2)\bar{\Delta}(z,M_1^2)
= \frac{1}{V}\sum_{p\neq 0}\frac{e^{ipx}}{(p^2+M_1^2)^2(p^2+M_2^2)}
\nonumber\\
= \frac{1}{(M_2^2-M_1^2)^2}
\left(\bar{\Delta}(x,M_2^2)-\bar{\Delta}(x,M_1^2)\right)
-\left.\frac{1}{M_2^2-M_1^2}\partial_{M^2}\bar{\Delta}(x,M^2)\right|_{M^2=M_1^2},
\end{eqnarray} 
whose degenerate limit, $M_1^2=M_2^2$, becomes
$(\partial_{M^2})^2\bar{\Delta}(x,M^2)|_{M^2=M_1^2}$.
We also need
\begin{eqnarray}
\int d^4 y\, d^4z\, \bar{\Delta}(x-y,M_1^2)\bar{G}(y-z,M_2^2,M_3^2)\bar{\Delta}(z,M_1^2)
\hspace{2in}\nonumber\\
= \frac
{\bar{G}(x,M_1^2,M_1^2)+\bar{G}(x,M_2^2,M_3^2)
-\bar{G}(x,M_1^2,M_3^2)-\bar{G}(x,M_1^2,M_2^2)
}{(M_2^2-M_1^2)(M_3^2-M_1^2)},
\end{eqnarray} 
which becomes in the limit $M_2^2=M_3^2$,
\begin{eqnarray}
= \frac
{\bar{G}(x,M_1^2,M_1^2)+\bar{G}(x,M_2^2,M_2^2)
-2\bar{G}(x,M_1^2,M_2^2)
}{(M_2^2-M_1^2)^2}.
\end{eqnarray} 

For the disconnected part, we compute
\begin{eqnarray}
\label{eq:DD4dint}
\int d^4 x \bar{\Delta}(x,M_1^2)\bar{\Delta}(x,M_2^2)
&=& \frac{1}{M_2^2-M_1^2}
\left(\bar{\Delta}(0,M_1^2)-\bar{\Delta}(0,M_2^2)\right),
\end{eqnarray}
and
\begin{eqnarray}
\label{eq:DG4dint}
\int d^4 x \bar{\Delta}(x,M_1^2)\bar{G}(x,M_2^2,M_3^2)
&=& \frac{1}{2}\left[
\frac{1}{M_1^2-M_2^2}
\left(\bar{G}(0,M_2^2,M_3^2)-\bar{G}(0,M_1^2,M_3^2)\right)
\right.\nonumber\\&&\left.
+\frac{1}{M_1^2-M_3^2}
\left(\bar{G}(0,M_2^2,M_3^2)-\bar{G}(0,M_1^2,M_2^2)
\right)\right],
\end{eqnarray}
of which divergent part is treated with
the dimensional regularization as usual.
\section{$\xi$  contraction in  the pseudoscalar correlator}
\label{app:PP00etc}

Here we summarize the $\xi$ contractions 
in 
 $\langle P(x)P(0)\rangle^{00}$, 
$\langle P(x)P(0)\rangle^{10}$, $\langle P(x)P(0)\rangle^{20}$ 
and $\langle P(x)P(0)\rangle^{01}$.

The first leading contribution is given by
\begin{eqnarray}
\langle P(x)P(0)\rangle^{00}
&=& -\frac{\Sigma^2(Z_M^{12}Z_F^{12})^4}{4}
\left\langle ([U_0]_{12}-[U_0^\dagger]_{21})
([U_0]_{21}-[U_0^\dagger]_{12})
\right.\nonumber\\&&\left.\hspace{1.2in}
+\frac{1}{2}([U_0]_{12}-[U_0^\dagger]_{21})^2
+\frac{1}{2}([U_0]_{21}-[U_0^\dagger]_{12})^2
\right\rangle_{U_0}
\nonumber\\&&
-\frac{\Sigma^2}{8}\left(\Delta Z^{\Sigma}_{11}-\Delta Z^{\Sigma}_{22}\right)
\langle ([U_0]_{12}-[U_0^\dagger]_{21})^2-([U_0]_{21}-[U_0^\dagger]_{12})^2\rangle_{U_0}
\nonumber\\&&
+\frac{\Sigma^2}{2F^2}(Z_F^{12}(Z_M^{12})^2)^2
\bar{\Delta}(x,M_{12}^{\prime 2})
\langle ([U_0]_{11}+[U_0^\dagger]_{22})([U_0]_{22}+[U_0^\dagger]_{11})\rangle_{U_0}
\nonumber\\&&
+\frac{\Sigma^2}{2F^2}\sum_{j\neq 1}
\bar{\Delta}(x,M_{j2}^2)
\langle [U_0]_{1j}[U_0^\dagger]_{j1}\rangle_{U_0}
\nonumber\\&&
+\frac{\Sigma^2}{2F^2}\sum_{i\neq 2}
\bar{\Delta}(x,M_{i1}^2)
\langle [U_0]_{2i}[U_0^\dagger]_{i2}\rangle_{U_0}
\nonumber\\&&
-\frac{\Sigma^2}{2F^2}\bar{G}(x,M_{11}^2,M_{22}^2)
\langle ([U_0]_{12}+[U_0^\dagger]_{21})([U_0]_{21}+[U_0^\dagger]_{12})\rangle_{U_0}
\nonumber\\&&
+\frac{\Sigma^2}{4F^2}
\bar{\Delta}(x,M_{22}^2)
\langle ([U_0]_{12})^2+([U_0^\dagger]_{21})^2\rangle_{U_0}
\nonumber\\&&
+\frac{\Sigma^2}{4F^2}
\bar{\Delta}(x,M_{11}^2)
\langle ([U_0]_{21})^2+([U_0^\dagger]_{12})^2\rangle_{U_0}
\nonumber\\&&
-\frac{\Sigma^2}{4F^2}
\bar{G}(x,M_{22}^2,M_{22}^2)
\langle ([U_0]_{12}+[U_0^\dagger]_{21})^2\rangle_{U_0}
\nonumber\\&&
-\frac{\Sigma^2}{4F^2}
\bar{G}(x,M_{11}^2,M_{11}^2)
\langle ([U_0]_{21}+[U_0^\dagger]_{12})^2\rangle_{U_0},
\end{eqnarray} 
where we have used
\[
\Sigma_{\rm eff}^2\left(1-\Delta Z^{\Sigma}_{22}+\frac{16 L_8}{F^2}M_{12}^2\right)
\left(1-\Delta Z^{\Sigma}_{11}+\frac{16 L_8}{F^2}M_{12}^2\right) =\Sigma^2(Z_M^{12}Z_F^{12})^4.
\]

Next we calculate the ${\cal O}(\mathcal{S}_{I}^{(1)})$ contribution.
In this NLO part, we can set $Z_\xi^{ij}=Z_F^{ij}=Z_M^{ij}=1$.
Note that $\xi$ contractions have to be all {\it connected}
since the self-contraction is not allowed in the
NSC vertex in $\mathcal{S}_{I}^{(1)}$.

Using a notation given in Eq.~(\ref{eq:R}) and the integration formulas
given in Appendix~\ref{app:xi}, we obtain
\begin{eqnarray}
\langle P(x)P(0)\rangle^{10} &=& 
\frac{\Sigma^2}{2F^2}
2\left\langle 
[\mathcal{R}]_{11}
+[\mathcal{R}]_{22}
\right\rangle_{U_0}
\left.\left(\frac{\Sigma}{F^2}\partial_{M^2}\right) 
\bar{\Delta}(x, M^2)\right|_{M^2=M_{12}^2}
\nonumber\\&&
-\frac{\Sigma^2}{2F^2}\sum^{N_f}_{j\neq 1}
\frac{\langle
[\mathcal{R}]_{j1}[U_0]_{1j}+
[\mathcal{R}]_{1j}[U^\dagger_0]_{j1}
\rangle_{U_0}}{m_j-m_1}
\left(\bar{\Delta}(x, M_{12}^2)-\bar{\Delta}(x, M_{2j}^2)\right)
\nonumber\\&&
-\frac{\Sigma^2}{2F^2}\sum^{N_f}_{i\neq 2}
\frac{\langle
[\mathcal{R}]_{i2}[U_0]_{2i}+
[\mathcal{R}]_{2i}[U^\dagger_0]_{i2}
\rangle_{U_0}}{m_i-m_2}
\left(\bar{\Delta}(x, M_{12}^2)-\bar{\Delta}(x, M_{1i}^2)\right)
\nonumber\\&&
-\frac{\Sigma^2}{2F^2}
\frac{\langle 
[\mathcal{R}]_{12}
[U^\dagger_0]_{12}+[U_0]_{21}[\mathcal{R}]_{21}
\rangle_{U_0}}{m_1-m_2}
\left(\bar{\Delta}(x, M_{12}^2)-\bar{\Delta}(x, M_{11}^2)\right)
\nonumber\\&&
-\frac{\Sigma^2}{2F^2}
\frac{\langle 
[\mathcal{R}]_{21}
[U^\dagger_0]_{21}+[U_0]_{12}[\mathcal{R}]_{12}
\rangle_{U_0}}{m_2-m_1}
\left(\bar{\Delta}(x, M_{12}^2)-\bar{\Delta}(x, M_{22}^2)\right)
\nonumber\\&&
+\frac{\Sigma^2}{2F^2}
\frac{\langle 
[\mathcal{R}]_{12}
([U_0]_{21}+[U_0^\dagger]_{12})+([U_0]_{21}+[U_0^\dagger]_{12})[\mathcal{R}]_{21}
\rangle_{U_0}}{m_1-m_2}
\nonumber\\&&\hspace{2in}
\times\left(
\bar{G}(x, M_{11}^2,M_{22}^2)-\bar{G}(x, M_{11}^2,M_{11}^2)\right)
\nonumber\\&&
+\frac{\Sigma^2}{2F^2}
\frac{\langle 
[\mathcal{R}]_{21}
([U_0]_{12}+[U_0^\dagger]_{21})+([U_0]_{12}+[U_0^\dagger]_{21})[\mathcal{R}]_{12}
\rangle_{U_0}}{m_2-m_1}
\nonumber\\&&\hspace{2in}
\times\left(
\bar{G}(x, M_{11}^2,M_{22}^2)-\bar{G}(x, M_{22}^2,M_{22}^2)\right).
\nonumber\\
\end{eqnarray}

For the ${\cal O}\left((\mathcal{S}_{I}^{(1)})^2\right)$ contribution
we have both {\it connected} and {\it disconnected} parts.
Note that we can set $Z_\xi^{ij}=Z_F^{ij}=Z_M^{ij}=1$ here, too.

The connected part (noted by the subscript ``$con$'') is given by
\begin{eqnarray}
\langle P(x)P(0)\rangle^{20}_{con}
&=& \left.\frac{\Sigma^2}{2F^2}\right[
\nonumber\\&&
-\left(
\sum_{j\neq 1}\frac{\langle 
[\mathcal{R}]_{1j}[\mathcal{R}]_{j1}
\rangle_{U_0}}{(m_j-m_1)}
+\sum_{i\neq 2}\frac{\langle 
[\mathcal{R}]_{2i}[\mathcal{R}]_{i2}
\rangle_{U_0}}{(m_i-m_2)}
\right)
\left.\left(\frac{\Sigma}{F^2}\partial_{M^2}\right)
\bar{\Delta}(x,M^2)\right|_{M^2=M_{12}^2}
\nonumber\\&&
-\sum_{j\neq 1}\frac{\langle 
[\mathcal{R}]_{1j}[\mathcal{R}]_{j1}
\rangle_{U_0}}{(m_j-m_1)^2}
\left(\bar{\Delta}(x,M_{12}^2)-\bar{\Delta}(x,M_{2j}^2)\right)
\nonumber\\&&
-\sum_{i\neq 2}\frac{\langle 
[\mathcal{R}]_{2i}[\mathcal{R}]_{i2}
\rangle_{U_0}}{(m_i-m_2)^2}
\left(\bar{\Delta}(x,M_{12}^2)-\bar{\Delta}(x,M_{1i}^2)\right)
\nonumber\\&&
+\frac{\langle 
([\mathcal{R}]_{12})^2+
([\mathcal{R}]_{21})^2\rangle_{U_0}}{2(m_2-m_1)^2}
\left(\bar{\Delta}(x,M_{11}^2)+\bar{\Delta}(x,M_{22}^2)
-2\bar{\Delta}(x,M_{12}^2)\right)
\nonumber\\&&
+\left.\frac{\langle 
2[\mathcal{R}]_{12}[\mathcal{R}]_{21}
+
([\mathcal{R}]_{12})^2+
([\mathcal{R}]_{21})^2\rangle_{U_0}}{(m_1-m_2)^2}
\bar{A}(x,M_{11}^2,M_{22}^2)\right].
\end{eqnarray}

For the disconnected contribution, we first calculate
\begin{eqnarray}
\frac{1}{2}\langle(\mathcal{S}^{(1)}_{I})^2\rangle_\xi
&=& -\frac{\Sigma^2V}{8 F^2}\sum_{i \neq j}
\frac{2\mathcal{R}_{ij}\mathcal{R}_{ji}}{M_{ii}^2-M_{jj}^2}
\left(\Delta Z_{ii}^{\Sigma}-\Delta Z_{jj}^{\Sigma}\right),
\end{eqnarray}
using Eqs.~(\ref{eq:DD4dint}) and (\ref{eq:DG4dint}) in Appendix~\ref{app:xi}.
Then we obtain (noted by the subscript ``$dis$'')
\begin{eqnarray}
\langle P(x)P(0)\rangle^{20}_{dis} &=&
\frac{1}{2}\left[\langle\mathcal{\alpha}(U_0)
\langle(\mathcal{S}^{(1)}_{I})^2\rangle_\xi\rangle_{U_0}
-\langle\mathcal{\alpha}(U_0)\rangle_{U_0}
\langle(\mathcal{S}^{(1)}_{I})^2\rangle_{\xi, U_0}\right]
\nonumber\\&&
+\frac{1}{2}\left[\langle\mathcal{\beta}(U_0)
\langle(\mathcal{S}^{(1)}_{I})^2\rangle_\xi\rangle_{U_0}
-\langle\mathcal{\beta}(U_0)\rangle_{U_0}
\langle(\mathcal{S}^{(1)}_{I})^2\rangle_{\xi, U_0}\right]\bar{\Delta}(x,M_{12}^2),
\end{eqnarray}
where
\begin{eqnarray}
\alpha(U_0) &\equiv& -\frac{\Sigma^2}{4}
\left[
([U_0]_{12}-[U_0^\dagger]_{21})
([U_0]_{21}-[U_0^\dagger]_{12})
\right.\nonumber\\&&\hspace{1.2in}\left.
+\frac{1}{2}([U_0]_{12}-[U_0^\dagger]_{21})^2
+\frac{1}{2}([U_0]_{21}-[U_0^\dagger]_{12})^2
\right],\\
\beta (U_0) &\equiv& \frac{\Sigma^2}{2F^2}
([U_0]_{11}+[U_0^\dagger]_{22})([U_0]_{22}+[U_0^\dagger]_{11}).
\end{eqnarray}

Since $\Delta Z_{ii}^{\Sigma}$ rapidly decreases
as the mass $m_i$ reaches the $\epsilon$ regime,
the contribution is important only deeply inside the
$p$ regime. Therefore, we can perturbatively 
perform this part of the $U_0$ integral in advance.
Using the technique presented in Appendix~\ref{app:U0pert},
the calculation is given by
\begin{eqnarray}
\langle\mathcal{\alpha}(U_0)
\mathcal{R}_{ij}\mathcal{R}_{ji}\rangle_{U_0}
-\langle\mathcal{\alpha}(U_0)\rangle_{U_0}
\langle \mathcal{R}_{ij}\mathcal{R}_{ji}\rangle_{U_0}
&=& \frac{4\Sigma^2(m_1-m_2)^2}{(\mu_1+\mu_2)^2}
(\delta_{i2}\delta_{j1}+\delta_{j2}\delta_{i1})+{\cal O}(p^9),\\
\langle\mathcal{\beta}(U_0)
\mathcal{R}_{ij}\mathcal{R}_{ji}\rangle_{U_0}
-\langle\mathcal{\beta}(U_0)\rangle_{U_0}
\langle \mathcal{R}_{ij}\mathcal{R}_{ji}\rangle_{U_0}
&=& {\cal O}(p^{10}),
\end{eqnarray}
where $\mu_i =m_i \Sigma V$ and we obtain
\begin{eqnarray}
\langle P(x)P(0)\rangle^{20}_{dis} &=&
- \frac{\Sigma^2(\mu_1-\mu_2)}{(\mu_1+\mu_2)^2}
\left\langle \frac{[U_0+U_0^\dagger]_{11}}{2}
+\frac{[U_0+U_0^\dagger]_{22}}{2}\right\rangle_{U_0}
\left(\frac{1}{2}\Delta Z_{11}^{\Sigma}
-\frac{1}{2}\Delta Z_{22}^{\Sigma}\right), \nonumber\\
\end{eqnarray}
where we have used
\begin{eqnarray}
2&=& \left\langle \frac{[U_0+U_0^\dagger]_{11}}{2}
+\frac{[U_0+U_0^\dagger]_{22}}{2}\right\rangle_{U_0} +{\cal O}(p^2)
\end{eqnarray}
for the later convenience.

Finally let us calculate the ${\cal O}(\mathcal{S}_I^{(2)})$ contribution.
As in the calculation above, using the technique in
Appendix~\ref{app:U0pert}, we obtain
\begin{eqnarray}
\langle P(x)P(0)\rangle^{01} &=&
- \frac{\Sigma^2}{(\mu_1+\mu_2)^2}
\left\langle \frac{[U_0+U_0^\dagger]_{11}}{2}
+\frac{[U_0+U_0^\dagger]_{22}}{2}\right\rangle_{U_0}
\nonumber\\&&\times
\left[
\mu_1\left(-\Delta Z_{11}^{\Sigma}
+\frac{16L_8}{F^2}M_{12}^2\right)
+\mu_2\left(-\Delta Z_{22}^{\Sigma}
+\frac{16L_8}{F^2}M_{12}^2\right)\right].
\end{eqnarray}
Here we note
\begin{eqnarray}
\langle P(x)P(0)\rangle^{20}_{dis}+\langle P(x)P(0)\rangle^{01}
\hspace{-2in}\nonumber\\ 
&=& \frac{\Sigma^2}{(\mu_1+\mu_2)}
\left\langle \frac{[U_0+U_0^\dagger]_{11}}{2}
+\frac{[U_0+U_0^\dagger]_{22}}{2}\right\rangle_{U_0}
\left(\frac{\Sigma_{\rm eff}}{\Sigma}-(Z_M^{12}Z_F^{12})^2\right).
\end{eqnarray}
In order to obtain the final expression in Eq.~(\ref{eq:PPd3x}),
we use
\begin{eqnarray}
-\frac{\Sigma^2}{4}(Z_M^{12}Z_F^{12})^4 \mathcal{C}^{0a}
+\frac{\Sigma^2}{\mu_1+\mu_2}
\left(\frac{\Sigma_{\rm eff}}{\Sigma}-(Z_M^{12}Z_F^{12})^2\right)
\mathcal{C}^{0b}
&=& \Sigma^2(Z_M^{12}Z_F^{12})^2
\frac{\mathcal{S}^{\rm eff}_1+\mathcal{S}^{\rm eff}_2}{\mu_1+\mu_2},
\end{eqnarray}
neglecting the higher order contributions.

\section{$U_0$ integrals }
\label{app:U0}

The zero-mode $U_0$ integrals of various matrix elements
have been calculated in Ref.~\cite{Damgaard:2007ep}.
Here we summarize the results 
in our notation for this paper.
\begin{eqnarray}
\frac{1}{2}\left\langle [U_0]_{vv}-[U_0^\dagger]_{vv}\right\rangle_{U_0}
&=& -\frac{Q}{\mu_v},\\
\frac{1}{4}\left\langle 
\left([U_0]_{vv}-[U_0^\dagger]_{vv}\right)^2\right\rangle_{U_0}
&=& -\frac{\mathcal{S}_v}{\mu_v}+\frac{Q^2}{\mu_v^2},\\
\frac{1}{4}\left\langle 
\left([U_0]_{v_1v_1}-[U_0^\dagger]_{v_1v_1}\right)
\left([U_0]_{v_2v_2}-[U_0^\dagger]_{v_2v_2}\right)\right\rangle_{U_0}
&=& \frac{Q^2}{\mu_{v_1}\mu_{v_2}},\\
\frac{1}{4}\left\langle 
\left([U_0]_{v_1v_2}\pm [U_0^\dagger]_{v_2v_1}\right)^2\right\rangle_{U_0}
= \frac{1}{4}\left\langle 
\left([U_0]_{v_2v_1}\pm [U_0^\dagger]_{v_1v_2}\right)^2\right\rangle_{U_0}
&=& \frac{\pm 1}{\mu_{v_1}^2-\mu_{v_2}^2}\left(
\mu_{v_1}\mathcal{S}_{v_1}-\mu_{v_2}\mathcal{S}_{v_2}\right),
\nonumber\\\\
\frac{1}{4}\left\langle 
\left([U_0]_{v_1v_2}\pm [U_0^\dagger]_{v_2v_1}\right)
\left([U_0]_{v_2v_1}\pm [U_0^\dagger]_{v_1v_2}\right)\right\rangle_{U_0}
&=& \frac{1}{\mu_{v_1}^2-\mu_{v_2}^2}\left(
\mu_{v_2}\mathcal{S}_{v_1}-\mu_{v_1}\mathcal{S}_{v_2}\right).
\nonumber\\
\end{eqnarray}
Here it is useful to define
\begin{eqnarray}
\delta_i \mathcal{S}_j &\equiv&  
\lim_{N_f+N\to N_f} \frac{\partial}{\partial \mu_{i}} \mathcal{S}_j,
\end{eqnarray}
or more explicitly,
\begin{eqnarray}
\delta_i \mathcal{S}_j=\left\{
\begin{array}{ll}  
\displaystyle\lim_{\mu_{b_1}\to \mu_{i}, \mu_{b_2}\to \mu_{j}}
\frac{\partial}{\partial \mu_{i}}
\frac{\partial}{\partial \mu_{j}} 
\ln \mathcal{Z}^Q_{2,2+N_f}(\mu_{b_1},\mu_{b_2}, \mu_{i},\mu_{j},\{\mu_{sea}\})
 & \mbox{($i\neq j$)},\\
\displaystyle\lim_{\mu_{b}\to \mu_{i}}
\frac{\partial^2}{\partial \mu_{i}^2}
\ln \mathcal{Z}^Q_{1,1+N_f}(\mu_{b}, \mu_{i},\{\mu_{sea}\})
 & \mbox{($i=j$)}.
\end{array}
\right.
\end{eqnarray}
Note that the partial quenching is performed {\it after}
the differentiation.
Then $\mathcal{D}$'s can be expressed as
\begin{eqnarray}
\mathcal{D}_i &=& \delta_i \mathcal{S}_i+\mathcal{S}_i^2,\\
\mathcal{D}_{ij} &=& \delta_i \mathcal{S}_j+\mathcal{S}_i\mathcal{S}_j
= \delta_j \mathcal{S}_i+\mathcal{S}_i\mathcal{S}_j.
\end{eqnarray}
We note
\begin{eqnarray}
m_i(\mathcal{S}_i-1) &\sim&  {\cal O}(p^4),\\
m_j m_i\delta_j\mathcal{S}_i &\sim&  {\cal O}(p^8),
\end{eqnarray}
which is useful to simplify our results.

We also note that 
$\mathcal{D}_{vv}$ 
(or $\mathcal{D}_{12}$ in the degenerate case $m_1=m_2=m_v$)
can be written in a simpler form than 
the original definition.
Introducing simplified notations
for the zero-mode partition functions:
\begin{eqnarray}
\mathcal{Z}_0 &=& \mathcal{Z}_{0,N_f}^Q(\{\mu_{sea}\}),\\
\mathcal{Z}_1(\mu_b| \mu_v) &=& 
\mathcal{Z}_{1,1+N_f}^Q(\mu_b,\mu_v,\{\mu_{sea}\}),\\
\mathcal{Z}_2(\mu_{b1},\mu_{b2}| \mu_{v1},\mu_{v2}) &=& 
\mathcal{Z}_{2,2+N_f}^Q(\mu_{b1},\mu_{b2},\mu_{v1},\mu_{v2},\{\mu_{sea}\}),
\end{eqnarray}
and noting that these partition functions satisfy
\begin{eqnarray}
\lim_{\mu_b \to \mu_v}\mathcal{Z}_1(\mu_b| \mu_v) &=& \mathcal{Z}_0,\\
\lim_{\mu_{b2}\to \mu_{v2}}
\mathcal{Z}_2(\mu_{b1},\mu_{b2}| \mu_{v1},\mu_{v2}) &=& 
\mathcal{Z}_1(\mu_{b1}| \mu_{v1}),\\
\lim_{\mu_{b2}\to \mu_{v1}}
\mathcal{Z}_2(\mu_{b1},\mu_{b2}| \mu_{v1},\mu_{v2}) &=& 
\mathcal{Z}_1(\mu_{b1}| \mu_{v2}),
\end{eqnarray}
it is easy to show 
\begin{eqnarray}
\left.\left(\frac{\partial}{\partial \mu_{bi}}
+\frac{\partial}{\partial \mu_{vi}}\right)
\mathcal{Z}_2(\mu_{b1},\mu_{b2}| \mu_{v1},\mu_{v2})
\right|_{\mu_{bi}=\mu_{vi}}=0
\end{eqnarray}
for any $i$.
We then obtain
\begin{eqnarray}
\mathcal{D}_{vv}
&=& - \left.\frac{1}{\mathcal{Z}_0}
\frac{\partial }{\partial \mu_{b}}\frac{\partial}{\partial \mu_{v}}
\mathcal{Z}_1(\mu_{b}| \mu_{v})
\right|_{\mu_{b}=\mu_v},
\end{eqnarray}
which is used to obtain expressions in Eqs.~(\ref{eq:Dvv 2-flavor}) 
and Eq.~(\ref{eq:Dvv 2+1-flavor}).

\section{$U_0$ integrals in the $p$ regime}
\label{app:U0pert}

In our calculation, we sometimes encounters
a situation that the zero-mode integrals are
needed only in the perturbative $p$ regime.
It is not impossible to nonperturbatively 
perform the zero-mode integrals even in such 
cases, but it is more convenient to go back
to the perturbative analysis to obtain the final results
in a simple form.

Let us start with an expansion of the $U_0$ field:
\begin{eqnarray}
U_0 &=& \exp\left(i\frac{\sqrt{2}\xi_0}{F}\right)
=1+\frac{i\sqrt{2}\xi_0}{F}-\frac{1}{F^2}\xi_0^2+\cdots ,
\end{eqnarray} 
and give a Feynman rule for $\xi_0$ 
\begin{eqnarray}
\langle [\xi_0]_{ij}[\xi_0]_{kl} \rangle &=&
\delta_{il}\delta_{jk}\frac{1}{M_{ij}^2 V}.
\end{eqnarray}
Note that it reproduces the ordinary propagator
in the $p$ expansion
together with $\bar{\Delta}(x,M_{ij}^2)$.
It is here important to note that $\xi_0$
is an element not of $SU(N)$ but of $U(N)$ 
Lie algebra and there is no diagonal contribution like 
non-zero mode $\xi$ has\footnote{
This argument is subtle for the summation over topology whose NNLO contribution
produces $\langle Q^2 \rangle /\mu_i\mu_j =1/\mu_i\mu_j(\sum_f 1/\mu_f)$,
which comes from the diagonal contribution,
$\langle [\xi_0]_{ii}[\xi_0]_{jj} \rangle$.
Fortunately, however, only
off-diagonal contributions are needed in the calculation of this paper,
and we can therefore ignore this subtlety.
}. 
Then we can calculate the zero-mode integrals 
in the $p$ regime as
\begin{eqnarray}
\langle [U_0]_{ij}[U_0]_{kl}\rangle_{U_0}
&=& - \delta_{il}\delta_{jk}\frac{2}{\mu_i+\mu_j}+{\cal O}(p^3),\\
\langle [U_0]_{ij}[U^\dagger_0]_{kl}\rangle_{U_0}
&=& + \delta_{il}\delta_{jk}\frac{2}{\mu_i+\mu_j}+{\cal O}(p^3),
\end{eqnarray}
\begin{eqnarray}
\frac{\left\langle [U_0]_{ij}[U_0]_{ji}
[U_0+U_0^\dagger]_{kk}
\right\rangle_{U_0}}{2}
-\frac{\left\langle [U_0]_{ij}[U_0]_{ji}\right\rangle_{U_0}
\left\langle
[U_0+U_0^\dagger]_{kk}
\right\rangle_{U_0}}{2}
&=& 2(\delta_{ik}+\delta_{jk})\left(\frac{1}{\mu_i+\mu_j}\right)^2,
\nonumber\\\\
\frac{\left\langle [U_0]_{ij}[U_0^\dagger]_{ji}
[U_0+U_0^\dagger]_{kk}
\right\rangle_{U_0}}{2}
-\frac{\left\langle [U_0]_{ij}[U_0^\dagger]_{ji}\right\rangle_{U_0}
\left\langle
[U_0+U_0^\dagger]_{kk}
\right\rangle_{U_0}}{2}
&=& -2(\delta_{ik}+\delta_{jk})\left(\frac{1}{\mu_i+\mu_j}\right)^2.
\nonumber\\
\end{eqnarray}
These results can be, of course, confirmed
by directly performing the exact group integrals
and then taking the asymptotic expansion in large $m_i \Sigma V$'s.
\section{Axialvector-Pseudoscalar correlator in the pure
$\epsilon$ regime}
\label {app:APepsilon}

In this appendix we present the
axialvector-pseudoscalar correlator
in the $\epsilon$ regime, which is, to our knowledge,
not found in the literature.

Since $M^2\sim {\cal O}(\epsilon^4)$ is deep inside the $\epsilon$ regime,
we can neglect the meson mass in the $Z$ factors: let us remove 
the superscripts and use notations such as $Z_M$, $Z_F$. We also note
$\Sigma_{\rm eff}= \Sigma Z_M^2 Z_F^2$
and $\Delta Z_{22}^{\Sigma}=0$ to NLO in the $\epsilon$ regime.

The source terms are then simplified as
\begin{eqnarray}
P^{12}(x) &=& i\frac{\Sigma_{\rm eff}}{2}
\left([U_0]_{12}-[U_0^\dagger]_{21}\right)
\nonumber\\&&
-\frac{\Sigma}{\sqrt{2}F}\sum_{i,j}
\xi_{ij}(x)\left([U_0]_{1i}\delta_{j2}+\delta_{1i}[U_0^\dagger]_{j2}\right)
Z_\xi Z_F(Z_M)^2\nonumber\\
&&-i\frac{\Sigma}{2F^2}\sum_{i,j}
[\xi^2(x)]_{ij}^{NSC}\left([U_0]_{1i}\delta_{j2}
-\delta_{1i}[U_0^\dagger]_{j2}\right),
\\
P^{21}(x) &=& (1\leftrightarrow 2),
\end{eqnarray}
and the axialvector sources can be similarly written as
\begin{eqnarray}
A_0^{12}(x) &=& -\frac{F}{\sqrt{2}}
\sum_{i,j}
[\partial_0 \xi(x)]_{ji}
\left([U^\dagger_0]_{i2}[U_0]_{1j}+\delta_{i2}\delta_{1j}\right)
Z_\xi Z_F
\nonumber\\&&
+\frac{i}{2}\sum_{i,j}
[\partial_0 \xi\xi-\xi\partial_0\xi]^{NSC}_{ji}(x)
\left([U_0^\dagger]_{i2}[U_0]_{1j}-\delta_{i2}\delta_{1j}\right),\\
A_0^{21}(x) &=& (1\leftrightarrow 2).
\end{eqnarray}

Note that the mass term is now an NLO contribution,
which can be treated as a perturbative interaction term
and one can omit the mass in the Feynman rule for $\xi$:
\begin{eqnarray}
\langle \xi_{ij}(x)\xi_{kl}(y)\rangle_\xi
&=&\delta_{il}\delta_{jk}\bar{\Delta}(x-y, 0)
-\delta_{ij}\delta_{kl}\bar{G}(x-y,0,0),
\end{eqnarray}

We therefore replace $\mathcal{S}_{I}^{(1)}$ by
\begin{eqnarray}
\mathcal{S}_I &\equiv& 
\frac{\Sigma}{2F^2}\int d^4 x
{\rm Tr}\left[\mathcal{R}^\prime[\xi^2(x)]^{NSC}\right],
\end{eqnarray}
where
\begin{eqnarray}
\mathcal{R}^\prime &\equiv& \mathcal{M}^\dagger U_0+
U_0^\dagger\mathcal{M}+\frac{N_f}{\Sigma V}{\bf 1}.
\end{eqnarray}
Since $\mathcal{S}_I \sim {\cal O}(\epsilon^2)$,
it is sufficient to calculate
\begin{eqnarray}
\langle A_0(x)P(0)\rangle &=&
\frac{1}{2}\left[
\langle A^{12}_0(x)P^{21}(0)+ A^{12}_0(x)P^{12}(0)\rangle^{00}
+\langle A^{12}_0(x)P^{21}(0)+ A^{12}_0(x)P^{12}(0)\rangle^{10}
\right]
\nonumber\\&&
+(1 \leftrightarrow 2).
\end{eqnarray}

Noting
$\left\langle [\partial_0\xi\xi
-\xi\partial_0 \xi]^{NSC}_{ji}(x)[\xi^2]^{NSC}_{kl}(0)\right\rangle_\xi=0$,
and (see Ref.~\cite{Bernardoni:2008ei})
\begin{eqnarray}
\langle[U_0\mathcal{M}U_0]_{11}\rangle_{U_0}
&=& m_1 -\frac{2}{\Sigma V}(N_f+Q)\langle[U_0]_{11}\rangle_{U_0},\\
\langle[U^\dagger_0\mathcal{M}U^\dagger_0]_{11}\rangle_{U_0}
&=& m_1 -\frac{2}{\Sigma V}(N_f-Q)\langle[U^\dagger_0]_{11}\rangle_{U_0},
\end{eqnarray}
\begin{eqnarray}
\left\langle
\left([U_0]_{12}+[U_0^\dagger]_{12}\right)
\left([U_0]_{21}+[U_0^\dagger]_{21}\right)\right\rangle_{U_0}
&\nonumber\\
& \hspace{-2.5in}= \frac{1}{4}\left\langle 
2\left([U_0]_{12}+[U_0^\dagger]_{21}\right)
\left([U_0]_{21}+[U_0^\dagger]_{12}\right)
+
2\left([U_0]_{12}-[U_0^\dagger]_{21}\right)
\left([U_0]_{21}-[U_0^\dagger]_{12}\right)
\right.\nonumber\\
&\hspace{-2.5in} \left.+\left([U_0]_{12}+[U_0^\dagger]_{21}\right)^2
+\left([U_0]_{21}+[U_0^\dagger]_{12}\right)^2
-\left([U_0]_{12}-[U_0^\dagger]_{21}\right)^2
-\left([U_0]_{21}-[U_0^\dagger]_{12}\right)^2
\right\rangle,\nonumber\\
\end{eqnarray}
and using the integration formulas in Appendix~\ref{app:xi},
we obtain the correlator,
\begin{eqnarray}
\langle A_0(x)P(0)\rangle 
&=&  \Sigma_{\rm eff}
\left(1+\mathcal{D}_{12}^{\rm eff}+\frac{Q^2}{\mu^{\rm eff}_1\mu^{\rm eff}_2}\right)
\partial_0 \bar{\Delta}(x,M_{12}^2)
\nonumber\\&&
+\Sigma_{\rm eff}
\left[\mathcal{S}_1^{\rm eff}+\mathcal{S}_2^{\rm eff}
-\left(1+\mathcal{D}_{12}^{\rm eff}+\frac{Q^2}{\mu^{\rm eff}_1\mu^{\rm eff}_2}\right)\right]
\partial_0 \bar{\Delta}(x,0)
\nonumber\\&&
-\left. M_{12}^2\Sigma \frac{\mathcal{S}_1-\mathcal{S}_2}{\mu_1-\mu_2}
\partial_{M^2} \partial_0 (\bar{G}(x,M^2,0)+\bar{G}(x,0,M^2))\right|_{M=0}.
\end{eqnarray}

\end{document}